\documentclass{aa}  

\usepackage{graphicx}
\usepackage{txfonts}
\begin{document}

   \title{Model-independent cosmic acceleration and type Ia supernovae
     intrinsic luminosity redshift dependence}

   \author{I. Tutusaus
          \inst{1}
          \and
          B. Lamine\inst{1}
         \and
         A. Blanchard\inst{1}}

   \institute{IRAP, Université de Toulouse, CNRS, CNES, UPS, (Toulouse), France\\
              \email{isaac.tutusaus@irap.omp.eu}
             }

   \date{Received --; accepted --}

 
  \abstract
   {The $\Lambda$CDM model is the current standard model in cosmology
     thanks to its ability to reproduce the observations. The first
     observational evidence for this model appeared roughly 20 years
     ago from the type Ia supernovae (SNIa) Hubble diagram from two
     different groups. However, there has been some debate in the
     literature concerning the statistical treatment of SNIa, and
     their ability to prove the cosmic acceleration.}
   {In this paper we relax the standard assumption that SNIa intrinsic
     luminosity is independent of the redshift, and we examine whether
     it may have an impact on our cosmological knowledge; more
     precisely, on the accelerated nature of the expansion
     of the Universe.}
   {In order to be as general as possible, we do not specify a given
     cosmological model, but we reconstruct the expansion rate of the
     Universe through a cubic spline interpolation fitting the
     observations of the different cosmological probes: SNIa, baryon
     acoustic oscillations (BAO), and the high-redshift information
     from the cosmic microwave background (CMB).}
   {We show that when SNIa intrinsic luminosity is not allowed to vary
   as a function of the redshift, cosmic acceleration is definitely
   proven in a model-independent approach. However, allowing for 
   a redshift dependence,
   a
   non-accelerated reconstruction of the expansion rate is able to
   fit, as well as $\Lambda$CDM, the combination of SNIa and BAO data, both treating the BAO standard ruler $r_d$
   as a free parameter (not entering on the physics governing the
   BAO), or adding the recently published prior from CMB observations. We further extend the
   analysis by including the CMB data. In this case we also consider a
   third way to combine the different probes by explicitly computing
   $r_d$ from the early Universe physics, and we show that a
   non-accelerated reconstruction is able to nicely fit this combination of
   low and high-redshift data. We also check
 that this reconstruction is compatible with the latest measurements of the growth rate of matter
 perturbations. We finally show that the value of the
 Hubble constant ($H_0$) predicted by this reconstruction is in tension with model-independent measurements.}
   {We present a model-independent reconstruction of a non-accelerated expansion rate
     of the Universe that is able to nicely fit all the main background
     cosmological probes. However, the predicted value of $H_0$ is in
     tension with recent direct measurements. Our analysis
     points out that a final, reliable, and consensual value for $H_0$
     would be critical to definitively prove the cosmic acceleration
     in a model-independent way.}


   \maketitle
%

\section{Introduction}
The cosmological concordance model ($\Lambda$CDM), mainly
composed of cold dark matter and dark energy,
provides an extremely precise description of the properties of our
Universe with very few parameters. However, recent
observations [\cite{CMB,SNIa,BAO1}] show that these components form
about 95\% of the energy content of the Universe, and their true
nature remain still unknown. The evidence for an accelerated expansion,
coming from the type Ia supernovae (SNIa) Hubble
diagram [\cite{Riess,Perlmutter}], was key to consider the
$\Lambda$CDM as the concordance model. But there has recently been a
debate in the literature wondering whether SNIa data alone, or
combined with other low-redshift cosmological probes, can prove
the accelerated expansion of the Universe
[\cite{Sarkar,Shariff,Rubin,Ringermacher,Tutusaus2,Dam,Lonappan,Haridasu,Lin,Lukovi}]. For
instance, the authors in \cite{Sarkar} claim that, allowing for the varying shape of
the light curve and extinction by dust, they find that SNIa data are
still quite consistent with a constant rate of expansion, while the
authors in \cite{Rubin} claim, redoing this analysis, a 11.2\,$\sigma$ confidence level for
acceleration with SNIa data alone in a flat universe.

In SNIa analyses it is usually assumed that two different SNIa in two
different galaxies with the same colour, stretch of the light-curve,
and host stellar mass, have on average the same intrinsic luminosity,
independently of the redshift. In this work we follow the approach of our previous analysis
[\cite{Tutusaus2}], and we relax this assumption by allowing these SNIa
to have different intrinsic luminosities as a function of the
redshift. Relaxing this redshift independence assumption has also been considered
in other analyses [\cite{Wright,Drell, SNIaevII, Nordin, Ferr2009}]. In
\cite{Tutusaus2} it was shown that a non-accelerated power law
cosmology was able to fit the main low-redshift cosmological probes:
SNIa, the baryon acoustic oscillations (BAO), the Hubble parameter as a
function of the redshift ($H(z)$), and measurements of the growth of
structures ($f\sigma_8(z)$), when some intrinsic luminosity redshift
dependence is allowed. Nevertheless, this specific power-law model is excluded when considering cosmic
microwave background (CMB) information (as it was shown in
[\cite{Tutusaus}]), and recently confirmed by the authors of
\cite{Riess15}, who showed
that such a model cannot fit the latest SNIa observations at $z>1$,
even when accounting for some luminosity evolution. In this paper we extend
our previous study with a model-independent analysis, and we include
the latest BAO observations as well as the complementary high-redshift
CMB data. In order to be as
general as possible, we follow the approach of \cite{Bernal} and
reconstruct the expansion rate at late times through a cubic spline
interpolation.

In Sect.\,\ref{sec2} we
present the different cosmological probes and the specific data sets considered in the
analysis. In Sect.\,\ref{sec3} we describe the methodology used to reconstruct
the expansion rate in a model-independent way. We provide the results of our study in Sect.\,\ref{sec4},
and we conclude in Sect.\,\ref{sec5}.

\section{Cosmological probes}\label{sec2}
In this section we present the different cosmological probes
considered in the analysis, as well as the specific data sets used.

\subsection{Type Ia supernovae}
Type Ia supernovae are considered standardizable candles and they are
useful to measure cosmological distances and break some
degeneracies present in other cosmological probes. The standard
observable used in SNIa measurements is the so-called distance
modulus,

\begin{equation}
\mu(z)=5\text{log}_{10}\left(\frac{H_0}{c}d_L(z)\right)\,,
\end{equation}
where $d_L(z)=(1+z)r(z)$ is the luminosity distance, and $r(z)$ the
comoving angular diameter distance.

The standardization of SNIa is based on empirical observation that
they form a homogeneous class of objects, whose variability can be
characterized by two parameters [\cite{SNIaparams}]: the time stretching of
the light curve ($X_1$) and the SNIa color at maximum brightness
($C$). If we assume that different SNIa with identical colour, shape,
and galactic environment have on average the same intrinsic luminosity
for all redshifts, the distance modulus can be expressed as

\begin{equation}
\mu_{\rm obs}=m_B^*-(M_B-\alpha X_1+\beta C)\,,
\end{equation}
where $m_B^*$ corresponds to the observed peak magnitude in the
$B$-band rest-frame, while $\alpha,\,\beta$ and $M_B$ are nuisance
parameters. Although the mechanism is not fully understood, it has
been shown [\cite{Sullivan,Johansson}] that both $\beta$ and $M_B$ depend on properties
of the host galaxy. In this work we use the joint light-curve
analysis from \cite{SNIa}, where the authors approximately correct for these
effects assuming that the absolute magnitude $M_B$ is related to the
stellar mass of the host galaxy, $M_{\rm stellar}$, by a simple step
function

\begin{equation}
M_B=\left\{
\begin{array}{ll}
M_B^1 & \text{if }M_{\rm stellar} < 10^{10} M_{\odot}\,,\\
M_B^1+\Delta_M & \text{otherwise}\,,
\end{array}
\right.
\end{equation}
where $M_B^1$ and $\Delta_M$ are two extra nuisance parameters. The
authors also discard the additional dependency of $\beta$ on the host stellar mass
because it does not have a significant impact on the cosmology.

Concerning the errors and the correlations of the measurements, we use
the full covariance matrix provided in \cite{SNIa}, where the authors
have considered several statistical and systematic uncertainties, such
as the error propagation of the light-curve fit uncertainties,
calibration, light-curve model, bias correction, mass step, dust
extinction, peculiar velocities, and contamination of non-type Ia
supernovae. This covariance matrix depends on the $\alpha$ and $\beta$
nuisance parameters, so when we sample the parameter space we
recompute the covariance matrix at each step.

Allowing for some redshift dependence on the SNIa intrinsic luminosity, the distance modulus can be expressed as

\begin{equation}
\mu_{\rm obs}=m_B^*-(M_B-\alpha X_1+\beta C+\Delta m_{\rm evo}(z))\,,
\end{equation}
where $\Delta m_{\rm evo}(z)$ stands for a nuisance term that accounts
for the intrinsic luminosity dependence as a function of the
redshift. 

Although the mechanism of SNIa detonation is well understood, the
difficulty of observing the system before becoming a SNIa leaves
enough uncertainty to wonder whether considering a luminosity
dependence with the redshift may have an effect on the cosmological
conclusions. A varying gravitational constant, or a fine structure
constant variation, could generate a luminosity dependence on the
redshift, but our approach here is just to consider a phenomenological
model to explore the degeneracy of SN distance-dependent effects and
the cosmological information. Different phenomenological models have been proposed for
$\Delta m_{\rm evo}(z)$ (see \cite{Tutusaus2} and references
therein). In this work we just consider Model B from \cite{Tutusaus2},
that has also been illustrated in \cite{Riess15}, where $\Delta m_{\rm
  evo}(z)=\epsilon z^{\delta}$. A lower $\delta$ power contribution
models a luminosity evolution dominant at low-redshift, while a higher
$\delta$ power contribution lead to a luminosity evolution dominating
at high-redshift. It is important to notice that $\delta$ must be
greater than $0$, in order not to be degenerate with $M_B^1$. When
sampling the parameter space we limit $\delta \in [0.2,2]$.

\subsection{Baryon acoustic oscillations}\label{BAOopts}
The baryon acoustic oscillations are the characteristic patterns
observed in the galaxy distribution of the large-scale structure of
the Universe. They are characterized by the length of a
standard ruler, $r_d$, and, in the standard cosmological model, they
are originated from sound waves propagating in the early Universe. The
BAO scale $r_d$ corresponds then to the comoving sound horizon at the redshift
of the baryon drag epoch,

\begin{equation}\label{eqrd}
r_d= r_s(z_{\rm drag})=\int_{z_{\rm drag}}^{\infty}\frac{c_s(z)\,\text{d}z}{H(z)}\,,
\end{equation}
where $z_{\rm drag}\approx 1060$ and $c_s(z)$ is the sound velocity as
a function of the redshift,

\begin{equation}
c_s(z)=\frac{c}{\sqrt{3(1+R_b(z))}}\,,\text{ with }R_b(z)=\frac{3\rho_b}{4\rho_{\gamma}}\,.
\end{equation}

In this expression $\rho_b$ stands for the baryon density while
$\rho_{\gamma}$ corresponds to the photon density. Their ratio can be
approximated [\cite{Rb}] by $R_b(z)=3.15\times 10^4 \Omega_b h^2\Theta_{2.7}^{-4}
(1+z)^{-1}$, with $\Theta_{2.7}=T_{\rm CMB}/2.7\,\text{K}$ and
$\Omega_b$ the baryon energy density parameter. In this work we fix
the temperature of the CMB to $T_{\rm CMB}=2.725\,$K [\cite{Fixsen}].

However, it is known that models differing from the standard
$\Lambda$CDM framework may have a value for $r_d$ that is not
compatible with $r_s(z_{\rm drag})$ [\cite{Verde}], and it has also been shown that
the computation of $r_d$ may have an effect on the trouble with the
Hubble constant $H_0$ [\cite{Bernal}]. Moreover, there has recently
been some analyses computing $r_d$ without any dependence on late-time
Universe assumptions [\cite{Verde2}]. Because of all this, in this work
we consider three different methods to include BAO data: either we
compute $r_d$ with equation (\ref{eqrd}), or we let it free, or we include
the prior $r_d=147.4\pm 0.7\,$Mpc from \cite{Verde2}.

We consider both isotropic and anisotropic measurements
of the BAO. The distance scale used for isotropic measurements is
given by

\begin{equation}
D_V(z)=\left(r^2(z)\frac{cz}{H(z)}\right)^{1/3}\,,
\end{equation}
while for the radial and transverse measurements the distance scales
are $r(z)$ and $H(z)$, respectively.

We use the isotropic measurements provided by 6dFGS at $z=0.106$
[\cite{BAO1}] and by
SDSS - MGS at $z=0.15$ [\cite{BAO2}], as well as the results from
WiggleZ at $z=0.44,0.6,0.73$ [\cite{WiggleZ}]. We also consider the anisotropic
final results of BOSS DR12
at $z=0.38, 0.51, 0.61$ [\cite{BOSSfinal}], and the new anisotropic
measurements from the eBOSS DR14 quasar sample [\cite{eBOSS1}] at $z=1.19,1.50,1.83$. These
results have been obtained by measuring the redshift space distortions
using the power spectrum monopole, quadrupole and hexadecapole. The
authors in \cite{eBOSS1} have shown that their results are completely
consistent with different methods used for analyzing the same data
[\cite{eBOSS2,eBOSS3}]. We finally consider the latest results from the
combination of the Ly-$\alpha$ forest auto-correlation function
[\cite{Lya1}] and
the Ly$\alpha$-quasar cross-correlation function [\cite{Lya2}] from the complete
BOSS survey at $z=2.4$. We take into account the covariance matrix
provided for
the measurements of WiggleZ, BOSS DR12, eBOSS DR14, we consider a
correlation coefficient of -0.38 for the Ly-$\alpha$ forest
measurements, and we assume measurements of different surveys to be
uncorrelated. In order to take into account the non-Gaussianity of the
BAO observable likelihoods far from the peak, we follow \cite{BAONG}
by replacing the usual $\Delta \chi^2_G=-2\text{ln}\mathcal{L}_G$ for
a Gaussian likelihood observable by

\begin{equation}
\Delta \chi^2=\frac{\Delta \chi^2_G}{\sqrt{1+\Delta \chi^4_G\left(\frac{S}{N}\right)^{-4}}}\,,
\end{equation}
where the ratio $S/N$ stands for the detection significance, in units
of $\sigma$, of the BAO feature. We consider a detection significance
of $2.4\,\sigma$ for 6dFGS, $2\,\sigma$ for SDSS-MGS and WiggleZ,
$9\,\sigma$ for BOSS DR12, $4\,\sigma$ for eBOSS DR14, and $5\,\sigma$
for the Ly-$\alpha$ forest.

\subsection{Cosmic microwave background}
The cosmic microwave background is an extremely powerful source of
information due to the high precision of modern data. Furthermore it
represents high-redshift data, complementing low-redshift probes. As
it was shown in \cite{Wang}, a significant part of the information
coming from the CMB can be compacted into a few numbers, the so-called
reduced parameters: the scaled distance to recombination $R$, the
angular scale of the sound horizon at recombination $\ell_a$, and the
reduced density parameter of baryons $\omega_b$. For a flat universe
their expressions are given by

\begin{align}
R&\equiv \sqrt{\Omega_m H_0^2}\int_0^{z_{\rm CMB}}
   \frac{\text{d}z}{H(z)}\,,\nonumber\\
\ell_a&\equiv\frac{\pi c}{r_s(z_{\rm CMB})}\int_0^{z_{\rm
        CMB}}\frac{\text{d}z}{H(z)}\,,\\
\omega_b&\equiv \Omega_b h^2\,,\nonumber
\end{align}
where $z_{\rm CMB}$ stands for the redshift of the last scattering
epoch. In this work we consider the data obtained from the Planck 2015
data release [\cite{CMBredparams}], where the compressed likelihood
parameters are obtained from the Planck temperature-temperature plus
the low-$\ell$ Planck temperature-polarization likelihoods. We
specifically consider the reduced parameters obtained when
marginalizing over the amplitude of the lensing power spectrum for the
lower values, since it leads to a more conservative approach, together
with their covariance matrix.

\section{Methodology}\label{sec3}
In this section we first remind the standard $\Lambda$CDM model and we
then present the reconstruction method used to obtain the expansion
rate as a function of the redshift. We give a detailed explanation of
how we introduce each cosmological probe in the analysis, and we
finally describe the method used to sample the parameter space.

\subsection{The $\Lambda$CDM model}
The flat $\Lambda$CDM model assumes a flat Robertson-Walker metric
together with Friedmann-Lema\^{i}tre dynamics, leading to the comoving
angular diameter distance,

\begin{equation}
r(z)=c\int_0^z\frac{\text{d}z'}{H(z')}\,,
\end{equation}
and the Friedmann-Lema\^{i}tre equation,

\begin{equation}\label{eqez}
E(z)^2\equiv\frac{H(z)^2}{H_0^2}=\Omega_r(1+z)^4+\Omega_m(1+z)^3+(1-\Omega_r-\Omega_m)\,,
\end{equation}
where $\Omega_m$ ($\Omega_r$) stands for the matter (radiation) energy
density parameter. We follow \cite{CMB} in computing the radiation
contribution as

\begin{equation}\label{rad1}
\Omega_r=\Omega_{\gamma}\left[1+N_{\rm eff}\frac{7}{8}\left(\frac{4}{11}\right)^{4/3}\right]\,,
\end{equation}
where $\Omega_{\gamma}$ corresponds to the photon contribution

\begin{equation}\label{rad2}
\Omega_{\gamma}=4\times 5.6704\times 10^{-8}\frac{T_{\rm
    CMB}^4}{c^3}\frac{8\pi G}{3H_0^2}\,.
\end{equation}

In this work we fix the effective number of neutrino-like relativistic
degrees of freedom to $N_{\rm eff}=3.04$. When we consider only SNIa
data, or SNIa combined with BAO data letting $r_d$ free, we fix the value of
$H_0$ for the radiation contribution on $\Lambda$CDM [see equations (\ref{rad1},\,\ref{rad2})] to $H_0=68\,$km$\,$s$^{-1}\,$Mpc$\,^{-1}$, since there is
no sensitivity to $H_0$ in these cases. However, $H_0$ is left free
for all the other cases and reconstructions in the rest of the
work. The remaining parameters when fitting $\Lambda$CDM to the data
are $\Omega_m$ and the corresponding nuisance parameters of the
cosmological probes considered (see Table \ref{table1}).

\subsection{Expansion rate reconstruction method}
We want our reconstruction to be as model-independent as
possible, and we impose a smooth and continuous expansion
rate. Several modelindependent reconstruction methods have been used in
the literature to reconstruct the dark energy equation of state
parameter, or even the Hubble parameter. Among them let us mention the principal component
analysis [\cite{PCA1,PCA2,PCA3,PCA4,PCA5}], the Gaussian processes [\cite{GP1,GP2,GP3,GP4,GP5,GP6}], or, very
recently, the Weighted Polynomial Regression method [\cite{GV}]. In
this work we follow the approach from \cite{Bernal}, reconstructing the
late-time expansion history by expressing $E(z)\equiv H(z)/H_0$ in
piece-wise natural cubic splines. When we consider SNIa data alone, $E(z)$ is specified by its values at
5 different ``knots'' in redshift:
$z=0.1,0.25,0.57,0.8,1.3$. Therefore, our reconstruction when
analyzing SNIa data considers the following set of parameters
$\{h_i,\alpha,\beta,M,\Delta_M,\epsilon,\delta\}$ with $h_i$ for $i\in
[1,5]$ being the 5 knots in redshift, $\alpha,\beta,M,\Delta_M$ the
standard SNIa nuisance parameters, and $\epsilon,\delta$ the SNIa
intrinsic luminosity evolution parameters.

When BAO data is added into the analysis we consider an extra knot in
our reconstruction at $z=2.4$. We follow two different approaches to
include the BAO measurements: first we consider the product $H_0r_d$
as a free parameter, and secondly we add information coming from the
early Universe through the prior on $r_d$ from \cite{Verde2},
$r_d=147.4\pm 0.7\,$Mpc. In the first case, the set of parameters
considered in our reconstruction of $E(z)$ is $\{h_i,\alpha,\beta,M,\Delta_M,H_0r_d,\epsilon,\delta\}$ with $h_i$ for $i\in
[1,6]$ being the 6 knots in redshift, while in the second case we
consider $H_0$ and $r_d$ separately
$\{h_i,\alpha,\beta,M,\Delta_M,H_0,r_d,\epsilon,\delta\}$.

When we finally add the reduced parameters for the CMB we need to
specify $E(z)$ up to early-times. In order to do this we add the
seventh knot at $z=2.7$ computed according to a matter dominated 
model (with flat Robertson-Walker metric and Friedmann-Lema\^{i}tre
dynamics) with free $H_0$ and $\Omega_m$ parameters [see equation (\ref{eqez})], and we extend the model up
to very high-redshift. The main idea
in this reconstruction is to start at early-times following a
matter dominated model (plus radiation and a negligible contribution
of dark energy through a cosmological constant) and, when we start to have low-redshift data and a
cosmological constant is still negligible with respect to the quantity
of matter present in the Universe, we reconstruct $E(z)$ through a
cubic spline interpolation; in this way we give our reconstruction the
freedom to choose the preferred expansion without specifying a
particular model for dark energy. When analyzing the data we
consider three different cases, depending on the way of introducing
the BAO measurements. First, we consider $r_d$ as a free parameter,
while, in a second place, we add the
prior on $r_d$ from \cite{Verde2}. In
both cases, the set of parameters that enters into the reconstruction
is given by $\{h_i,\alpha,\beta,M,\Delta_M,H_0,r_d,\Omega_m,z_{\rm
  CMB},\omega_b,\epsilon,\delta\}$, and we add the prior on $z_{\rm
  CMB}=1089.90\pm 0.23$ [\cite{CMB}]. As a last case we compute the
value of $r_d$ using equation (\ref{eqrd}). In this case the set of
parameters is given by $\{h_i,\alpha,\beta,M,\Delta_M,H_0,\Omega_m,z_{\rm
  CMB},\omega_b,z_{\rm drag},\epsilon,\delta\}$, and we add the prior
on $z_{\rm drag}=1059.68\pm 0.29$ [\cite{CMB}].

In order to test the degeneracy between a SNIa intrinsic luminosity
redshift dependence and cosmic acceleration, we consider different cases with
and without evolution, so $\epsilon$ and $\delta$ can be removed from
the analysis, and we also consider coasting reconstructions, in which
the universe has a late-time constant expansion rate. More
specifically, we fix the first 4 knots (3 for SNIa alone) such that $E(z)$ is equal to
$(1+z)$ at these points. See Table\,\ref{table1} for a summary of the
different cases considered and the cosmological and nuisance
parameters present in them.

\begin{table*}
  \caption{Summary of the cosmological probes and parameters present in the
    different cases considered. The $i$-index on $h_i$ goes from 1 to
    5 for SNIa data alone, while it goes up to 6 when BAO data is
    included. When working with coasting reconstructions we only
    consider the last two knots $h_i$.}\label{table1}
\begin{center}
\begin{tabular}[c]{llll}
Case & Cosmological probes &Cosmological parameters &Nuisance parameters\\\hline \hline
SNIa&SNIa&$h_i$&$\alpha,\beta,M,\Delta_M$\\[2mm]
SNIa+BAO free
  $H_0r_d$&SNIa+BAO&$h_i,H_0r_d$&$\alpha,\beta,M,\Delta_M$\\[2mm]
SNIa+ev+BAO free
  $H_0r_d$&SNIa+BAO&$h_i,H_0r_d$&$\alpha,\beta,M,\Delta_M,\epsilon,\delta$\\[2mm]
SNIa+BAO prior
  $r_d$&SNIa+BAO&$h_i,H_0,r_d$&$\alpha,\beta,M,\Delta_M$\\[2mm]
SNIa+ev+BAO prior
  $r_d$&SNIa+BAO&$h_i,H_0,r_d$&$\alpha,\beta,M,\Delta_M,\epsilon,\delta$\\[2mm]
SNIa+BAO free
  $r_d$+CMB&SNIa+BAO+CMB&$h_i,H_0,r_d,\Omega_m,\omega_b$&$\alpha,\beta,M,\Delta_M,z_{\rm
                                                         CMB}$\\[2mm]
SNIa+ev+BAO free
  $r_d$+CMB&SNIa+BAO+CMB&$h_i,H_0,r_d,\Omega_m,\omega_b$&$\alpha,\beta,M,\Delta_M,z_{\rm
                                                         CMB},\epsilon,\delta$\\[2mm]
SNIa+BAO prior
  $r_d$+CMB&SNIa+BAO+CMB&$h_i,H_0,r_d,\Omega_m,\omega_b$&$\alpha,\beta,M,\Delta_M,z_{\rm
                                                         CMB}$\\[2mm]
SNIa+ev+BAO prior
  $r_d$+CMB&SNIa+BAO+CMB&$h_i,H_0,r_d,\Omega_m,\omega_b$&$\alpha,\beta,M,\Delta_M,z_{\rm
                                                         CMB},\epsilon,\delta$\\[2mm]
SNIa+BAO compute
  $r_d$+CMB&SNIa+BAO+CMB&$h_i,H_0,\Omega_m,\omega_b$&$\alpha,\beta,M,\Delta_M,z_{\rm
                                                         CMB},z_{\rm
                                                         drag}$\\[2mm]
SNIa+ev+BAO compute
  $r_d$+CMB&SNIa+BAO+CMB&$h_i,H_0,\Omega_m,\omega_b$&$\alpha,\beta,M,\Delta_M,z_{\rm
                                                         CMB},z_{\rm
                                                         drag}\epsilon,\delta$
\end{tabular}
\end{center}
\end{table*}

\subsection{Fitting the data}
In order to reconstruct the expansion rate as a function of the
redshift, we fit the data minimizing the common $\chi^2$ function,

\begin{equation}
\chi^2=(\textbf{u}-\textbf{u}_{\rm
  data})^TC^{-1}(\textbf{u}-\textbf{u}_{\rm data})\,,
\end{equation}
where $\textbf{u}$ stands for the model prediction, while
$\textbf{u}_{\rm data}$ and $C$ hold for the observables and their
covariance matrix, respectively. We sample the parameter space to
minimize this function using the MIGRAD application from the
\texttt{iminuit} Python
package\footnote{\url{https://github.com/iminuit/iminuit}}, a Python
implementation of the former MINUIT Fortran code [\cite{minuit}],
conceived to find the minimum value of a multiparameter function and
analyze the shape of the function around the minimum. We use it to
extract the best-fit values for the parameters, as well as their
errors and the covariance matrix of the parameters.

We also compute the probability that a higher value for the $\chi^2$
occurs for a fit with $\nu=N-k$ degrees of freedom, where $N$ is the
number of data points and $k$ is the number of parameters,

\begin{equation}\label{pchi2}
P(\chi^2,\nu)=\frac{\Gamma\left(\frac{\nu}{2},\frac{\chi^2}{2}\right)}{\Gamma\left(\frac{\nu}{2}\right)}\,,
\end{equation}
where $\Gamma(t,x)$ is the upper incomplete gamma function and
$\Gamma(t)=\Gamma(t,0)$ the complete gamma function. This value
provides us with a goodness of fit statistic. A probability close to 1
indicates that it is likely to obtain higher $\chi^2$ values than the
minimum found, pointing to a good fit by the model. When we combine different probes, we minimize the sum of the
individual $\chi^2$ functions for each probe, i.e., we assume the
probes to be uncorrelated.

\section{Results}\label{sec4}
In this section we present the results of the reconstruction of the
expansion rate of the Universe as a function of the redshift for
different sets of cosmological probes: SNIa, SNIa combined with BAO,
and SNIa combined with both BAO and CMB data. We also comment on the
linear growth of structures measurements, and the importance of the
value of the Hubble constant, $H_0$, to draw conclusions on the accelerated expansion of the Universe.

\subsection{Case 1: SNIa}
\begin{sidewaystable}
  \caption{Best-fit values with the corresponding 1$\sigma$ error bars
  for the cosmological and nuisance parameters of the first case: SNIa
data. The values of $\Lambda$CDM are added as a reference. The reduced
$\chi^2$ and the probability $P(\chi^2,\nu)$ are also provided for the
different models.}\label{table2}
\begin{center}
\resizebox{\textwidth}{!}{
\begin{tabular}[c]{ll|llllllllll|ll}
Case & Model  & $h_1$ & $h_2$ & $h_3$ & $h_4$ & $h_5$ & $\Omega_m$ &
                                                                    $\alpha$
  & $\beta$ & $M$ & $\Delta_M$ &
                                                         $\chi^2/$d.o.f&$P(\chi^2,\nu)$\\[2mm]\hline \hline 
&$\Lambda$CDM& $-$  & $-$ & $-$ & $-$ & $-$ &
                                                                     $0.295\pm
                                                                     0.034$&
                                                                 $0.1412\pm
                                                                 0.0066$
  & $3.101\pm 0.081$ & $24.110\pm 0.023$ & $-0.070\pm 0.023$ &
                                                               682.89/735
                                               & 0.915\\[2mm]
SNIa&Splines& $1.041\pm 0.022$ & $1.141\pm 0.023$ & $1.344\pm
                                                        0.071$ &
                                                                 $1.46\pm
                                                                 0.13$
                                                          & $1.90\pm
                                                            0.90$ &
                                                                 $-$ &
                                                                 $0.1414\pm
                                                                 0.0066$
  & $3.106\pm 0.082$ & $24.110\pm 0.032$ & $-0.070\pm 0.023$ &
                                                               681.38/731
                                               & 0.905\\[2mm]
&CS (3 knots)& $-$ & $-$ & $-$ &
                                                                 $1.69\pm
                                                                 0.15$
                                                          & $1.27\pm
                                                            0.58$& $-$
 &
                                                                 $0.1385\pm
                                                                 0.0066$
  & $3.075\pm 0.081$ & $24.230\pm 0.017$ & $-0.077\pm 0.023$ &
                                                               717.60/734
                                                                       &
                                                                         0.661\\[2mm]
\end{tabular}
}
\end{center}
\end{sidewaystable}

\begin{figure}
\begin{center}
\includegraphics[scale=0.55]{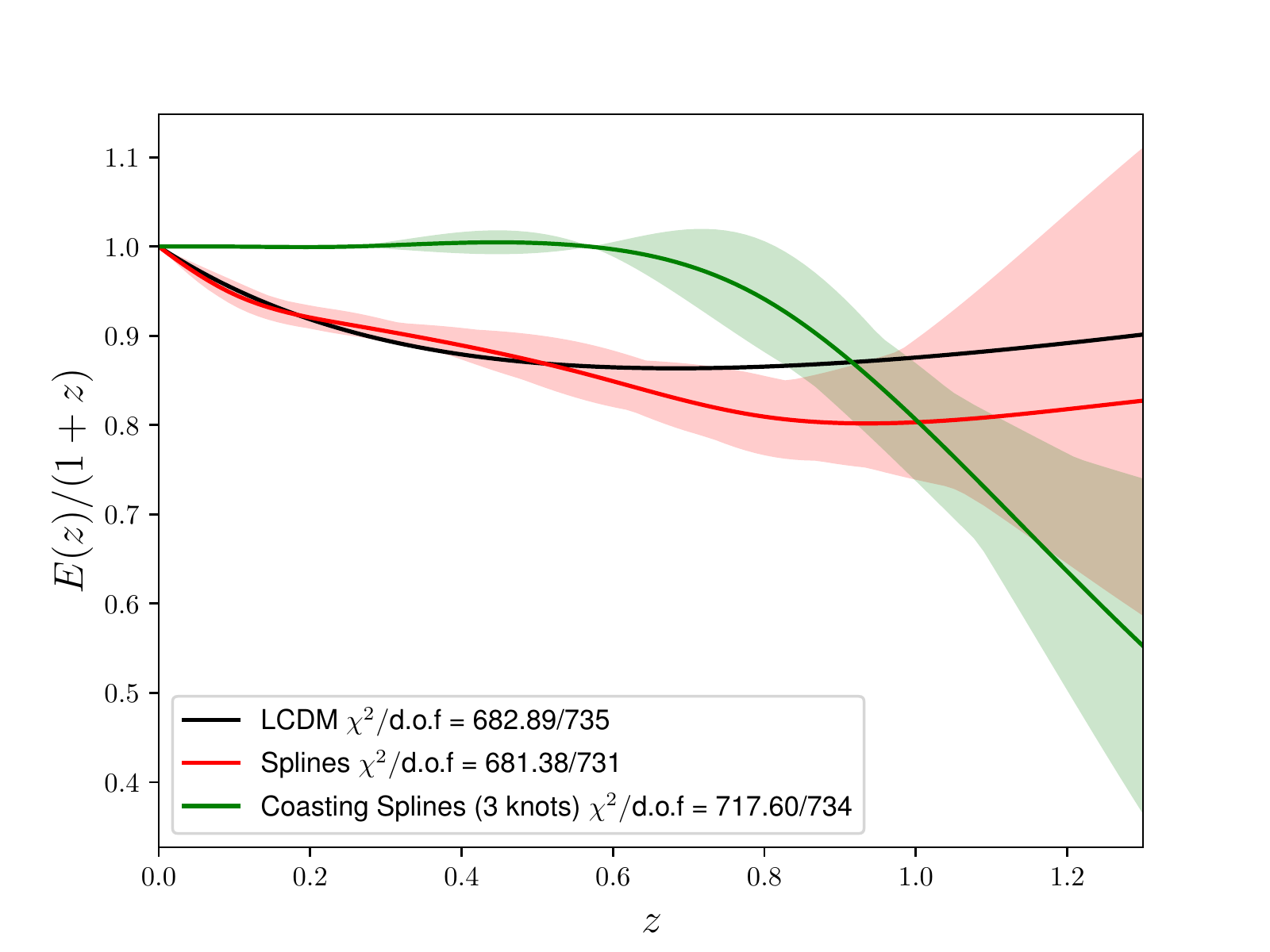}
\caption{Reconstruction of the expansion rate, $E(z)/(1+z)$, as a function of the
  redshift using SNIa data alone. The black line represents the
  $\Lambda$CDM model, while the red band shows the reconstruction with
  $\Delta \chi^2=1$ with respect to the best reconstruction (red
  line). The green band stands for the reconstruction of a coasting
  universe at low-redshift. See the text for the details of the
  reconstruction.}\label{fig1}
\end{center}
\end{figure}
We first start considering only SNIa data. We present this case as an
illustration of the reconstruction method used. The best-fit values
for the cosmological and nuisance parameters are presented in Table
\ref{table2} together with the 1$\sigma$ error bars, and the
reconstructions are shown in Fig. \ref{fig1}. We show three
different models in this case: the reconstruction through cubic
splines (red), the reconstruction for a coasting universe (labelled CS) at low-redshift
(fixing the first 3 knots - green), and $\Lambda$CDM as a reference (black). We do not
consider any SNIa luminosity evolution for the moment. In Table \ref{table2}
we also provide the reader with the ratio of the $\chi^2$ over the
number of degrees of freedom, and the probability $P(\chi^2,\nu)$ from
equation (\ref{pchi2}). In order to obtain the bands for the
reconstructions we generate 500 splines from an $N$-dimensional
Gaussian centered at the best-fit values and with the covariance
matrix obtained from the fit to the data. We further require each
spline to have a $\Delta\chi^2$ value smaller or equal than 1 with
respect to the best-fit reconstruction.

In Table \ref{table2} we can clearly see that all the SNIa nuisance
parameters values are compatible for the three models, and that a
coasting universe shows a lower expansion rate when we increase the
redshift with respect to the standard spline reconstruction. This is
confirmed from Fig. \ref{fig1} where we see that the expansion rate
drops above $z\approx 0.8$. We can also observe in this figure that the
bands increase when we increase the redshift, as expected, since there
are less data points in this region. Comparing the three models, we
observe that the spline reconstruction provides a slightly smaller
$\chi^2$ value (681.38) than $\Lambda$CDM (682.89), but the former has
many more parameters in the model, so the ability of these models to
fit the data is roughly the same, being slightly better for
$\Lambda$CDM ($P(\chi^2,\nu)=0.915$) than the spline reconstruction
($P(\chi^2,\nu)=0.905$). However, the $\chi^2$ value obtained for the
coasting reconstruction (717.60) is much larger than the previous
values, which also implies that this model is less able to perfectly
fit the data ($P(\chi^2,\nu)=0.661$). A detailed model comparison is
beyond the scope of this work, since we are interested in studying the
accelerated expansion of the Universe and the relation it may have
with SNIa luminosity, not in proposing a new cosmological model to
confront against $\Lambda$CDM. However, the coasting reconstruction
has a relative probability of exp($-\Delta\chi^2/2)\approx 1.4\times
10^{-6}$\,\%, showing that a coasting universe at low-redshift is
totally excluded, even using SNIa data alone, when SNIa intrinsic
luminosity is assumed to be redshift independent. Notice also that,
even if we ask the reconstruction to be non-accelerated at
low-redshift, it prefers to add some acceleration at earlier times
(above $z\approx 0.8$) than just having a constant velocity expansion.

\subsection{Case 2: SNIa+BAO}
\begin{sidewaystable}
  \caption{Best-fit values with the corresponding 1$\sigma$ error bars
  for the cosmological and nuisance parameters of the second case:
  SNIa and BAO data. The values of $\Lambda$CDM are added as a reference. The reduced
$\chi^2$ and the probability $P(\chi^2,\nu)$ are also provided for the
different models.}\label{table3}
\begin{center}
\resizebox{\textwidth}{!}{
\begin{tabular}[c]{ll|llllllllllllllll|ll}
Case & Model  & $h_1$ & $h_2$ & $h_3$ & $h_4$ & $h_5$ &
                                                                    $h_6$
  & $H_0r_d$ & $H_0$ & $r_d$ & $\Omega_m$ &
                                                                    $\alpha$
  & $\beta$ & $M$ & $\Delta_M$ & $\epsilon$ & $\delta$ &
                                                         $\chi^2/$d.o.f&$P(\chi^2,\nu)$\\[2mm]\hline \hline 
&$\Lambda$CDM& $-$  & $-$ & $-$ & $-$ & $-$ & $-$&$10120\pm 126$  & $-$ & $-$
                             &$0.292\pm 0.017$  &
                                                                 $0.1413\pm
                                                                 0.0066$
  & $3.102\pm 0.080$ & $24.110\pm 0.019$ & $-0.070\pm 0.023$ &
                                                               $-$ &
                                                                     $-$
                                                                & 
                                                               698.64/753
                                               & 0.922\\[2mm]
SNIa+BAO free $H_0r_d$&Splines& $1.050\pm 0.020$ & $1.133\pm 0.020$ & $1.385\pm
                                                        0.035$ &
                                                                 $1.591\pm
                                                                 0.074$
                                                          & $2.15\pm
                                                            0.14$ &
                                                                    $3.43\pm
                                                                    0.10$&$10040\pm
                                                                           174$
             & $-$ & $-$ & $-$ &
                                                                 $0.1410\pm
                                                                 0.0066$
  & $3.099\pm 0.081$ & $24.120\pm 0.030$ & $-0.070\pm 0.023$ &
                                                               
                                                                     $-$
                                                       & $-$ & 
                                                               696.46/748
                                               & 0.911\\[2mm]
&CS (4 knots)& $-$ & $-$ & $-$ &
                                                                 $-$
                                                          & $2.33\pm
                                                            0.15$&
                                                                   $3.768\pm
                                                                   0.098$
  & $9144\pm 69$ & $-$ & $-$ & $-$ &
                                                                 $0.1382\pm
                                                                 0.0066$
  & $3.073\pm 0.080$ & $24.230\pm 0.017$ & $-0.077\pm 0.023$ &
                                                               
                                                                     $-$
                                                       & $-$ & 
                                                               739.91/752
                                                                       &
                                                                         0.616\\[2mm]
  \hline

&$\Lambda$CDM& $-$  & $-$ & $-$ & $-$ & $-$ & $-$& $10120\pm
                                                               137$ &
                                                                      $-$
                                                                      &
                                                                        $-$
                                                                        &
                                                                          $0.292\pm
                                                                        0.019$&
                                                                              
                                                                 $0.1413\pm
                                                                 0.0066$
  & $3.103\pm 0.080$ & $24.110\pm 0.058$ & $-0.070\pm 0.023$ & 
                                                               
                                                                      $0.001\pm 0.080$ &
                                                                     $0.20\pm
                                                                                         0.19$ &
                                                               698.64/751
                                               & 0.914\\[2mm]
SNIa+ev+BAO free $H_0r_d$&Splines& $1.118\pm 0.044$ & $1.241\pm 0.065$ & $1.510\pm
                                                        0.078$ &
                                                                 $1.77\pm
                                                                 0.13$
                                                          & $2.34\pm
                                                            0.18$ &
                                                                    $3.70\pm
                                                                    0.19$&$9286\pm
                                                                           420$
             & $-$ & $-$ & $-$  &
                                                                 $0.1414\pm
                                                                 0.0066$
  & $3.101\pm 0.081$ & $23.94\pm 0.11$ & $-0.070\pm 0.023$ &
                                                               
                                                                     $0.39\pm
                                                                               0.22$
                                                       & $0.20\pm 0.16$ &
                                                               694.21/746
                                               & 0.912\\[2mm]
&CS (4 knots)& $-$ & $-$ & $-$ &
                                                                 $-$
                                                          & $2.38\pm
                                                            0.15$&
                                                                   $3.760\pm
                                                                   0.090$
  &$9140\pm 64$ & $-$ & $-$ & $-$ &
                                                                 $0.1416\pm
                                                                 0.0064$
  & $3.104\pm 0.081$ & $24.050\pm 0.081$ & $-0.070\pm 0.022$ & 
                                                               
                                                                     $0.322\pm
                                                                              0.075$
                                                       & $0.41\pm 0.20$ &
                                                               698.95/750
                                                                       &
                                                                         0.909\\[2mm]
  \hline

&$\Lambda$CDM& $-$  & $-$ & $-$ & $-$ & $-$ & $-$& $-$ &
                                                                      $68.66\pm
                                                         0.91$
                                                                      &
                                                                        $147.40\pm
                                                                        0.70$
                                                                        &
                                                                          $0.292\pm
                                                                        0.017$&
                                                                                
                                                                 $0.1413\pm
                                                                 0.0066$
  & $3.103\pm 0.080$ & $24.110\pm 0.019$ & $-0.070\pm 0.023$ & $-$ &
                                                                     $-$ &

                                                               698.64/753
                                               & 0.922\\[2mm]
SNIa+BAO prior $r_d$&Splines& $1.050\pm 0.020$ & $1.133\pm 0.019$ & $1.385\pm
                                                        0.032$ &
                                                                 $1.591\pm
                                                                 0.073$
                                                          & $2.15\pm
                                                            0.14$ &
                                                                    $3.43\pm
                                                                    0.10$&$-$
             & $68.1\pm 1.2$ & $147.40\pm 0.70$ & $-$ &
                                                                 $0.1411\pm
                                                                 0.0066$
  & $3.099\pm 0.081$ & $24.120\pm 0.029$ & $-0.070\pm 0.023$ & $-$ & $-$ &
                                                               
                                                               696.46/748
                                               & 0.911\\[2mm]
&CS (4 knots)& $-$ & $-$ & $-$ &
                                                                 $-$
                                                          & $2.33\pm
                                                            0.15$&
                                                                   $3.768\pm
                                                                   0.098$
  &$-$ & $62.04\pm 0.55$ & $147.40\pm 0.70$ & $-$ &
                                                                 $0.1382\pm
                                                                 0.0066$
  & $3.073\pm 0.080$ & $24.230\pm 0.017$ & $-0.077\pm 0.023$ & $-$ &
                                                                     $-$ &

                                                               739.91/752
                                                                       &
                                                                         0.616\\[2mm]
  \hline

&$\Lambda$CDM& $-$  & $-$ & $-$ & $-$ & $-$ & $-$& $-$ &
                                                                      $68.48\pm
                                                         0.97$
                                                                      &
                                                                        $147.40\pm
                                                                        0.70$
                                                                        &
                                                                          $0.296\pm
                                                                        0.019$&
                                                                                
                                                                 $0.1414\pm
                                                                 0.0066$
  & $3.105\pm 0.081$ & $24.110\pm 0.020$ & $-0.070\pm 0.023$ & 
                                                               
                                                                      $0.029\pm
                                                                           0.052$
                                                                             &
                                                                     $2.0\pm
                                                                               1.7$ &
                                                               698.35/751
                                               & 0.915\\[2mm]
SNIa+ev+BAO prior $r_d$&Splines& $1.048\pm 0.021$ & $1.149\pm 0.023$ & $1.410\pm
                                                        0.039$ &
                                                                 $1.671\pm
                                                                 0.095$
                                                          & $2.20\pm
                                                            0.15$ &
                                                                    $3.46\pm
                                                                    0.11$&$-$
             & $67.3\pm 1.3$ & $147.40\pm 0.70$ & $-$ &
                                                                 $0.1413\pm
                                                                 0.0066$
  & $3.101\pm 0.081$ & $24.110\pm 0.031$ & $-0.070\pm 0.023$ & 
                                                               
                                                                     $0.094\pm
                                                                           0.065$
                                                       & $2.0\pm 1.5$ &
                                                               694.37/746
                                               & 0.912\\[2mm]
&CS (4 knots)& $-$ & $-$ & $-$ &
                                                                 $-$
                                                          & $2.38\pm
                                                            0.16$&
                                                                   $3.759\pm
                                                                   0.097$
  &$-$ & $62.01\pm 0.55$ & $147.40\pm 0.70$ & $-$ &
                                                                 $0.1416\pm
                                                                 0.0066$
  & $3.104\pm 0.081$ & $24.050\pm 0.094$ & $-0.070\pm 0.023$ & 
                                                               
                                                                     $0.322\pm
                                                                           0.078$
                                                       & $0.41\pm 0.24$ &
                                                               698.95/750
                                                                       &
                                                                         0.909\\[2mm]

\end{tabular}
}
\end{center}
\end{sidewaystable}

\begin{figure*}
\begin{center}
\includegraphics[scale=0.55]{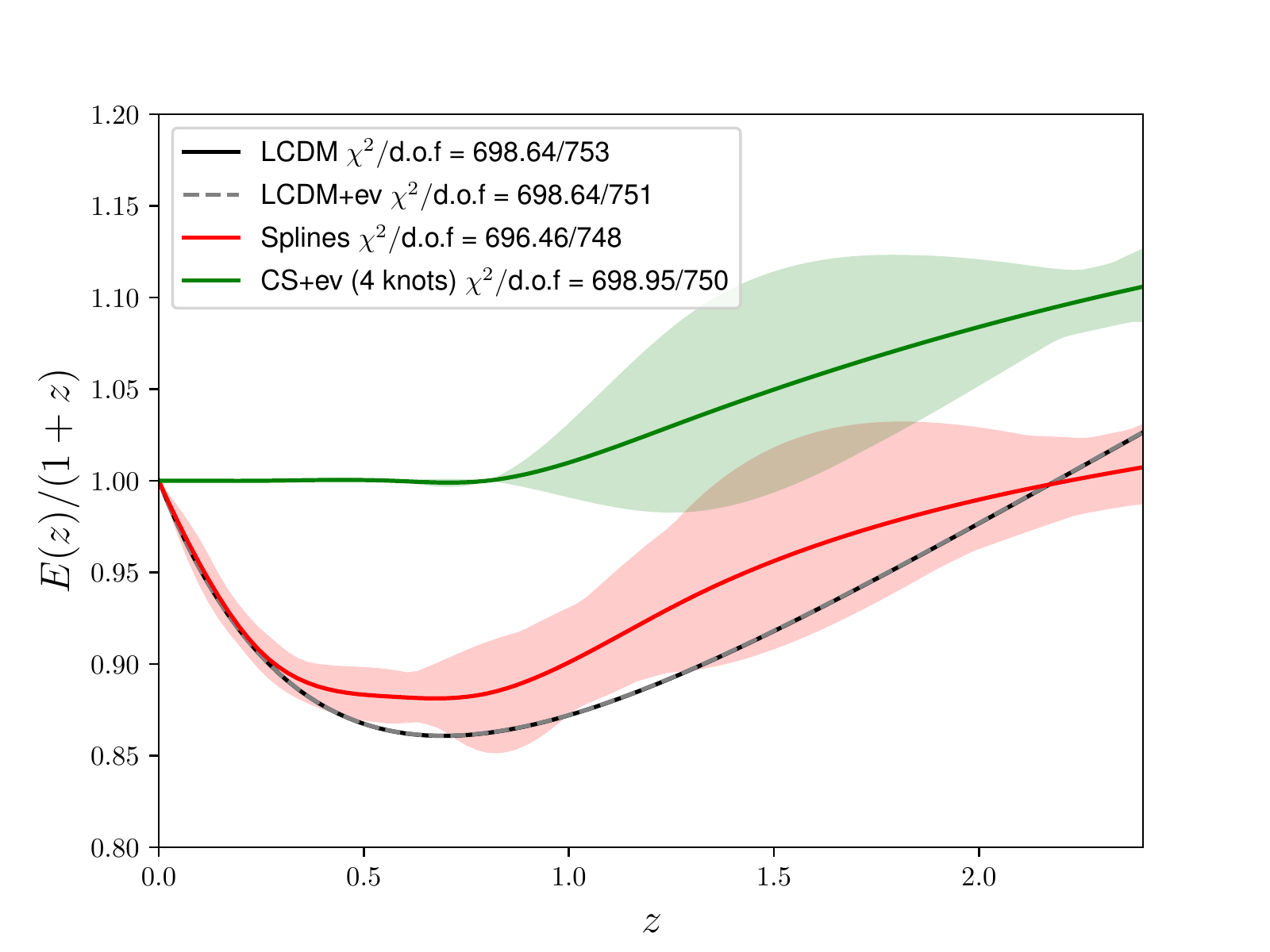}\,\includegraphics[scale=0.55]{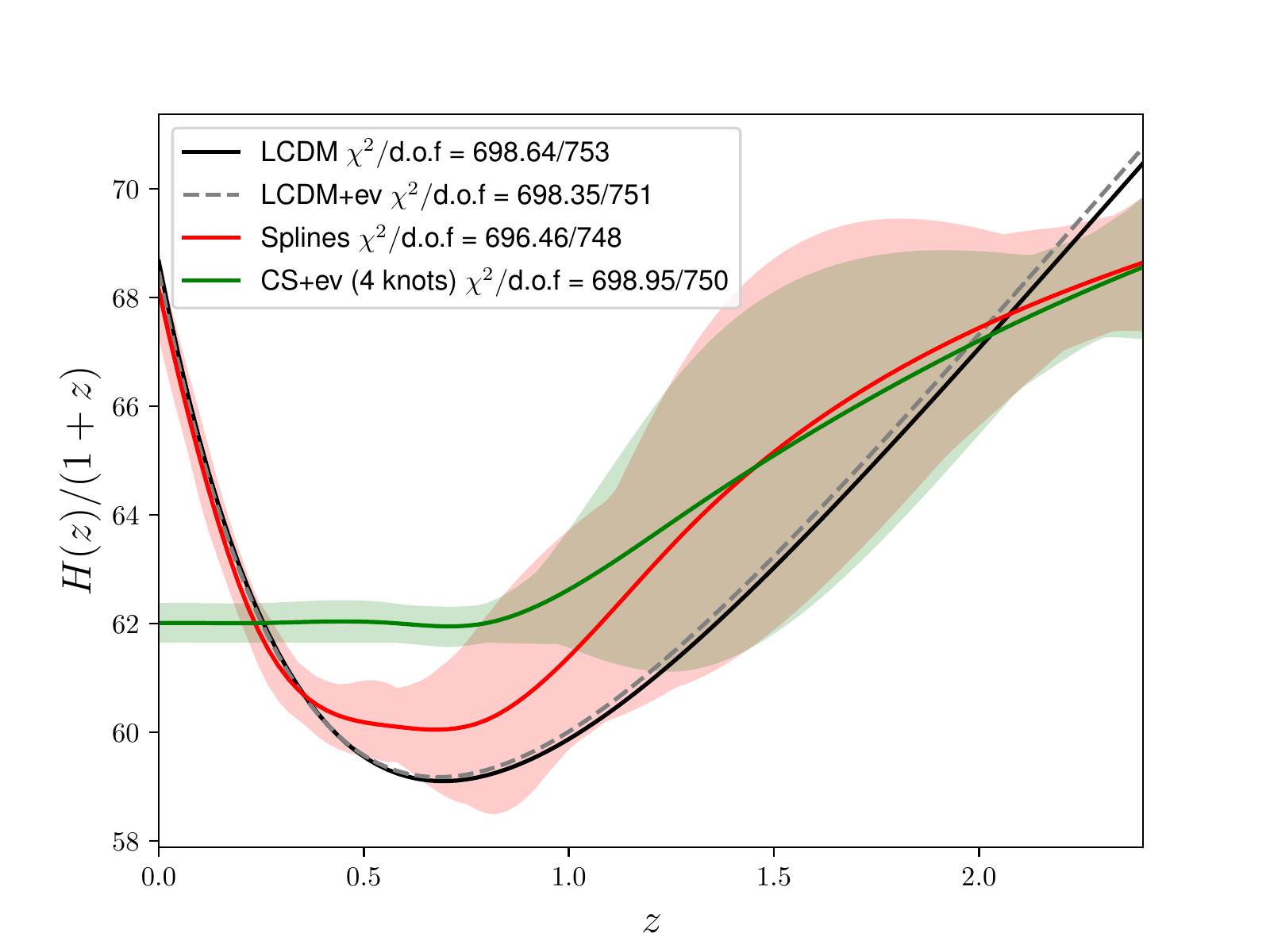}
\caption{Reconstruction of the expansion rate, $E(z)/(1+z)$ (left) and
  $H(z)/(1+z)$ (right), as a function of the
  redshift using the combination of SNIa and BAO data. In the left
  panel the data sets have been combined considering $H_0r_d$ a free
  parameter, while in the right panel a prior on $r_d$ has been added.
  In both panels the black and grey lines represent the
  $\Lambda$CDM model (without and with SNIa luminosity evolution, respectively), while the red band shows the reconstruction with
  $\Delta \chi^2=1$ with respect to the best reconstruction (red
  line). The green band stands for the reconstruction of a coasting
  universe at low-redshift when SNIa intrinsic luminosity is allowed
  to vary as a function of the redshift. See the text for the details
  of the reconstruction.}\label{fig2}
\end{center}
\end{figure*}

After having shown how the reconstruction method works, and having applied
it to SNIa data alone, we focus on the combination of SNIa and BAO
data. As it is shown in Table \ref{table1} we consider two different
ways to combine these data sets: we either let the product $H_0r_d$
free, or we add a prior on $r_d$. Since we consider the models with
and without SNIa intrinsic luminosity evolution, and we always add
$\Lambda$CDM as a reference, we finally have 4 different subcases
with the corresponding three models per subcase. The best-fit values
and errors for the parameters for all these cases are shown in Table
\ref{table3}.

Let us first focus on the case where $H_0r_d$ is treated as a free
parameter. As it was the case with SNIa data alone, all the SNIa
nuisance parameters have compatible values for the different models
considered. However, the coasting reconstruction now does not show a
reduced expansion rate at high-redshift (adding or not SNIa luminosity
evolution), due to the addition of the BAO data points above $z\sim
0.8$. We can also see that the value of $H_0r_d$ obtained from the
spline reconstruction is more or less compatible with the one obtained
with $\Lambda$CDM, but it is lower for the coasting reconstruction,
adding SNIa intrinsic luminosity or not. Concerning the ability of the
models to fit the data, the $\chi^2$ of the spline reconstruction is
always slightly smaller than the $\Lambda$CDM one (696.46 against
698.64, and 694.21 against 698.64 when we allow the SNIa luminosity to
vary). But as it was the case before, the probability of providing a
good fit is roughly the same for both models, being slightly better for
$\Lambda$CDM (0.911 against 0.922, and
0.912 against 0.914 when we account for evolution). It is also what
can be seen in the reconstruction plot on the left panel of
Fig. \ref{fig2}. With respect to the coasting reconstruction, we can
see in Table \ref{table3} that, when SNIa intrinsic luminosity is
allowed to vary, we obtain a $\chi^2$ value very close to the
$\Lambda$CDM one, thus giving also a good probability to correctly fit
the data (0.909 against 0.912, for the standard reconstruction, and
0.914, for $\Lambda$CDM). 

Let us now focus on the combination of SNIa and BAO data with a prior
on $r_d$ (two last rows of Table \ref{table3} and the
right panel of Fig. \ref{fig2}). It allows us to obtain a constraint on
$H_0$, so we represent in this case the expansion rate by
$H(z)/(1+z)$. All the best-fit values for the parameters are very
close to the previous case, with nearly the same $\chi^2$ values and
the same probabilities, since we have only added one data point and
one parameter in the analysis. As before, a coasting universe provides
a good fit to the data with a probability of 0.909 against 0.912, for
the standard spline reconstruction, and 0.915 for $\Lambda$CDM, when SNIa
luminosity is allowed to vary. The interesting result from these cases
is that the value found for $H_0$ for the spline reconstruction is
always smaller than the one obtained for $\Lambda$CDM, but still
compatible, while it is significantly smaller for the coasting
reconstruction, as it can be seen in the right plot of
Fig. \ref{fig2}. This is consistent with the lower value found for
$H_0r_d$ in the previous cases for the coasting reconstruction.

\subsection{Case 3: SNIa+BAO+CMB}
\begin{sidewaystable}
  \caption{Best-fit values with the corresponding 1$\sigma$ error bars
  for the cosmological and nuisance parameters of the third case:
  SNIa, BAO, and CMB data. The values of $\Lambda$CDM are added as a reference. The reduced
$\chi^2$ and the probability $P(\chi^2,\nu)$ are also provided for the
different models.}\label{table4}
\begin{center}
\resizebox{\textwidth}{!}{
\begin{tabular}[c]{ll|llllllllllllllllll|ll}
Case & Model  & $h_1$ & $h_2$ & $h_3$ & $h_4$ & $h_5$ &
                                                                    $h_6$
 & $H_0$ & $r_d$ & $\Omega_m$ & $10^2\omega_b$ &
                                                                    $\alpha$
  & $\beta$ & $M$ & $\Delta_M$ & $z_{\rm CMB}$ & $z_{\rm drag}$ & $\epsilon$ & $\delta$ &
                                                         $\chi^2/$d.o.f&$P(\chi^2,\nu)$\\[2mm]\hline \hline

&$\Lambda$CDM& $-$  & $-$ & $-$ & $-$ & $-$ & $-$ &
                                                                      $69.14\pm
                                                         0.93$
                                                                      &
                                                                        $146.2\pm
                                                                        1.4$
                                                                        &
                                                                          $0.293\pm
                                                                        0.011$&
                                                                                $2.264\pm
                                                                                0.025$
                                                       &
                                                                 $0.1412\pm
                                                                 0.0066$
  & $3.102\pm 0.080$ & $24.110\pm 0.018$ & $-0.070\pm 0.023$ &
                                                               $1090.00\pm 0.23$ &
                                                                     $-$ &
                                                               
                                                                      $-$
                                                                             &
                                                                     $-$ &
                                                               698.67/754
                                               & 0.926\\[2mm]
SNIa+BAO free $r_d$+CMB&Splines& $1.050\pm 0.020$ & $1.133\pm 0.020$ & $1.386\pm
                                                        0.034$ &
                                                                 $1.592\pm
                                                                 0.075$
                                                          & $2.14\pm
                                                            0.15$ &
                                                                    $3.45\pm
                                                                    0.10$
             & $68.4\pm 1.5$ & $146.7\pm 2.5$ & $0.300\pm 0.015$ &
                                                                   $2.263\pm 0.029$ &
                                                                 $0.1410\pm
                                                                 0.0066$
  & $3.099\pm 0.081$ & $24.120\pm 0.030$ & $-0.070\pm 0.023$ &
                                                               $1090.00\pm 0.23$ & $-$ &
                                                               
                                                                     $-$
                                                       & $-$ &
                                                               697.01/748
                                               & 0.909\\[2mm]
&CS (4 knots)& $-$ & $-$ & $-$ &
                                                                 $-$
                                                          & $2.29\pm
                                                            0.16$&
                                                                   $3.795\pm
                                                                   0.096$
  & $62.5\pm 1.1$ & $146.2\pm 2.5$ & $0.359\pm 0.014$ &
                                                             $2.264\pm 0.029$ &
                                                                 $0.1382\pm
                                                                 0.0066$
  & $3.073\pm 0.080$ & $24.230\pm 0.017$ & $-0.077\pm 0.023$ &
                                                               $1090.00\pm 0.23$ &
                                                                     $-$ &
                                                               
                                                                     $-$
                                                       & $-$ &
                                                               740.32/752
                                                                       &
                                                                         0.612\\[2mm]
  \hline

&$\Lambda$CDM& $-$  & $-$ & $-$ & $-$ & $-$ & $-$ &
                                                                      $69.11\pm
                                                         0.97$
                                                                      &
                                                                        $146.2\pm
                                                                        1.4$
                                                                        &
                                                                          $0.294\pm
                                                                        0.012$&
                                                                                $2.264\pm
                                                                                0.025$
                                                       &
                                                                 $0.1413\pm
                                                                 0.0066$
  & $3.103\pm 0.081$ & $24.110\pm 0.056$ & $-0.070\pm 0.023$ &
                                                               $1090.00\pm 0.23$ &
                                                                     $-$ &
                                                               
                                                                      $0.005\pm
                                                                           0.075$
                                                                             &
                                                                     $0.20\pm
                                                                               0.99$ &
                                                               698.66/752
                                               & 0.918\\[2mm]
SNIa+ev+BAO free $r_d$+CMB&Splines& $1.049\pm 0.021$ & $1.149\pm 0.023$ & $1.410\pm
                                                        0.039$ &
                                                                 $1.671\pm
                                                                 0.094$
                                                          & $2.18\pm
                                                            0.16$ &
                                                                    $3.48\pm
                                                                    0.10$
             & $67.3\pm 1.7$ & $147.5\pm 2.6$ & $0.310\pm 0.017$ &
                                                                   $2.263\pm 0.029$ &
                                                                 $0.1413\pm
                                                                 0.0066$
  & $3.101\pm 0.081$ & $24.110\pm 0.031$ & $-0.070\pm 0.023$ &
                                                               $1090.00\pm 0.23$ & $-$ &
                                                               
                                                                     $0.094\pm
                                                                                         0.064$
                                                       & $2.0\pm 1.6$ &
                                                               694.93/746
                                               & 0.909\\[2mm]
&CS (4 knots)& $-$ & $-$ & $-$ &
                                                                 $-$
                                                          & $2.36\pm
                                                            0.17$&
                                                                   $3.783\pm
                                                                   0.095$
  & $62.2\pm 1.1$ & $147.1\pm 2.6$ & $0.364\pm 0.015$ &
                                                             $2.263\pm 0.029$ &
                                                                 $0.1416\pm
                                                                 0.0066$
  & $3.104\pm 0.081$ & $24.050\pm 0.089$ & $-0.070\pm 0.023$ &
                                                               $1090.00\pm 0.23$ &
                                                                     $-$ &
                                                               
                                                                     $0.322\pm
                                                                           0.074$
                                                       & $0.41\pm 0.24$ &
                                                               699.44/750
                                                                       &
                                                                         0.906\\[2mm]
  \hline

&$\Lambda$CDM& $-$  & $-$ & $-$ & $-$ & $-$ & $-$ &
                                                                      $68.69\pm
                                                         0.72$
                                                                      &
                                                                        $147.20\pm
                                                                        0.63$
                                                                        &
                                                                          $0.2986\pm
                                                                        0.0090$&
                                                                                $2.255\pm
                                                                                0.021$
                                                       &
                                                                 $0.1412\pm
                                                                 0.0066$
  & $3.101\pm 0.080$ & $24.110\pm 0.018$ & $-0.070\pm 0.023$ &
                                                               $1090.00\pm 0.23$ &
                                                                     $-$ &
                                                               
                                                                      $-$
                                                                             &
                                                                     $-$ &
                                                               699.21/755
                                               & 0.927\\[2mm]
SNIa+BAO prior $r_d$+CMB&Splines& $1.050\pm 0.020$ & $1.133\pm 0.020$ & $1.385\pm
                                                        0.034$ &
                                                                 $1.595\pm
                                                                 0.074$
                                                          & $2.166\pm
                                                            0.096$ &
                                                                    $3.45\pm
                                                                    0.10$
             & $68.2\pm 1.2$ & $147.40\pm 0.68$ & $0.302\pm 0.011$ &
                                                                   $2.262\pm 0.028$ &
                                                                 $0.1410\pm
                                                                 0.0066$
  & $3.099\pm 0.081$ & $24.120\pm 0.030$ & $-0.070\pm 0.023$ &
                                                               $1090.00\pm 0.23$ & $-$ &
                                                               
                                                                     $-$
                                                       & $-$ &
                                                               697.07/749
                                               & 0.913\\[2mm]
&CS (4 knots)& $-$ & $-$ & $-$ &
                                                                 $-$
                                                          & $2.355\pm
                                                            0.089$&
                                                                   $3.792\pm
                                                                   0.095$
  & $62.09\pm 0.53$ & $147.30\pm 0.67$ & $0.3644\pm 0.0085$ &
                                                             $2.261\pm 0.028$ &
                                                                 $0.1382\pm
                                                                 0.0066$
  & $3.073\pm 0.080$ & $24.230\pm 0.017$ & $-0.077\pm 0.023$ &
                                                               $1090.00\pm 0.23$ &
                                                                     $-$ &
                                                               
                                                                     $-$
                                                       & $-$ &
                                                               740.51/753
                                                                       &
                                                                         0.620\\[2mm]
  \hline

&$\Lambda$CDM& $-$  & $-$ & $-$ & $-$ & $-$ & $-$ &
                                                                      $68.65\pm
                                                         0.75$
                                                                      &
                                                                        $147.20\pm
                                                                        0.63$
                                                                        &
                                                                          $0.2991\pm
                                                                        0.0093$&
                                                                                $2.254\pm
                                                                                0.022$
                                                       &
                                                                 $0.1413\pm
                                                                 0.0066$
  & $3.102\pm 0.081$ & $24.100\pm 0.056$ & $-0.070\pm 0.023$ &
                                                               $1090.00\pm 0.23$ &
                                                                     $-$ &
                                                               
                                                                      $0.015\pm
                                                                           0.073$
                                                                             &
                                                                     $0.2\pm
                                                                               1.7$ &
                                                               699.17/753
                                               & 0.920\\[2mm]
SNIa+ev+BAO prior $r_d$+CMB&Splines& $1.049\pm 0.021$ & $1.149\pm 0.023$ & $1.410\pm
                                                        0.038$ &
                                                                 $1.670\pm
                                                                 0.092$
                                                          & $2.175\pm
                                                            0.096$ &
                                                                    $3.48\pm
                                                                    0.10$
             & $67.3\pm 1.3$ & $147.40\pm 0.68$ & $0.310\pm 0.013$ &
                                                                   $2.264\pm 0.028$ &
                                                                 $0.1413\pm
                                                                 0.0066$
  & $3.101\pm 0.081$ & $24.110\pm 0.030$ & $-0.070\pm 0.023$ &
                                                               $1090.00\pm 0.23$ & $-$ &
                                                               
                                                                     $0.094\pm
                                                                                         0.063$
                                                       & $2.0\pm 1.6$ &
                                                               694.93/747
                                               & 0.913\\[2mm]
&CS (4 knots)& $-$ & $-$ & $-$ &
                                                                 $-$
                                                          & $2.374\pm
                                                            0.090$&
                                                                   $3.782\pm
                                                                   0.094$
  & $62.02\pm 0.53$ & $147.40\pm 0.67$ & $0.3650\pm 0.0085$ &
                                                             $2.263\pm 0.028$ &
                                                                 $0.1416\pm
                                                                 0.0066$
  & $3.104\pm 0.081$ & $24.050\pm 0.095$ & $-0.070\pm 0.023$ &
                                                               $1090.00\pm 0.23$ &
                                                                     $-$ &
                                                               
                                                                     $0.322\pm
                                                                           0.079$
                                                       & $0.41\pm 0.25$ &
                                                               699.46/751
                                                                       &
                                                                         0.911\\[2mm]
  \hline

&$\Lambda$CDM& $-$  & $-$ & $-$ & $-$ & $-$ & $-$ &
                                                                      $68.55\pm
                                                         0.59$
                                                                      &
                                                                        $-$
                                                                        &
                                                                          $0.3006\pm
                                                                        0.0071$&
                                                                                $2.254\pm
                                                                                0.022$
                                                       &
                                                                 $0.1411\pm
                                                                 0.0066$
  & $3.100\pm 0.080$ & $24.110\pm 0.018$ & $-0.070\pm 0.023$ &
                                                               $1090.00\pm 0.23$ &
                                                                     $1060.00\pm
                                                                                   0.29$ &
                                                               
                                                                      $-$
                                                                             &
                                                                     $-$ &
                                                               699.31/755
                                               & 0.927\\[2mm]
SNIa+BAO compute $r_d$+CMB&Splines& $1.050\pm 0.020$ & $1.133\pm 0.020$ & $1.385\pm
                                                        0.034$ &
                                                                 $1.595\pm
                                                                 0.074$
                                                          & $2.171\pm
                                                            0.096$ &
                                                                    $3.45\pm
                                                                    0.10$
             & $68.1\pm 1.2$ & $-$ & $0.303\pm 0.011$ &
                                                                   $2.261\pm 0.028$ &
                                                                 $0.1410\pm
                                                                 0.0066$
  & $3.099\pm 0.081$ & $24.120\pm 0.030$ & $-0.070\pm 0.023$ &
                                                               $1090.00\pm
                                                               0.23$ &
                                                                       $1060.00\pm 0.29$ &
                                                               
                                                                     $-$
                                                       & $-$ &
                                                               697.09/749
                                               & 0.912\\[2mm]
&CS (4 knots)& $-$ & $-$ & $-$ &
                                                                 $-$
                                                          & $2.362\pm
                                                            0.087$&
                                                                   $3.792\pm
                                                                   0.094$
  & $62.04\pm 0.47$ & $-$ & $0.3655\pm 0.0063$ &
                                                             $2.260\pm 0.027$ &
                                                                 $0.1382\pm
                                                                 0.0065$
  & $3.073\pm 0.078$ & $24.230\pm 0.017$ & $-0.077\pm 0.023$ &
                                                               $1090.00\pm 0.22$ &
                                                                     $1060.00\pm
                                                                                   0.29$ &
                                                               
                                                                     $-$
                                                       & $-$ &
                                                               740.55/753
                                                                       &
                                                                         0.620\\[2mm]
  \hline

&$\Lambda$CDM& $-$  & $-$ & $-$ & $-$ & $-$ & $-$ &
                                                                      $68.51\pm
                                                         0.60$
                                                                      &
                                                                        $-$
                                                                        &
                                                                          $0.3010\pm
                                                                        0.0073$&
                                                                                $2.253\pm
                                                                                0.022$
                                                       &
                                                                 $0.1413\pm
                                                                 0.0066$
  & $3.103\pm 0.081$ & $24.100\pm 0.056$ & $-0.070\pm 0.023$ &
                                                               $1090.00\pm 0.23$ &
                                                                     $1060.00\pm
                                                                                   0.29$ &
                                                               
                                                                      $0.019\pm
                                                                                           0.073$
                                                                             &
                                                                     $0.2\pm
                                                                               1.8$ &
                                                               699.24/753
                                               & 0.920\\[2mm]
SNIa+ev+BAO compute $r_d$+CMB&Splines& $1.049\pm 0.020$ & $1.149\pm 0.023$ & $1.410\pm
                                                        0.039$ &
                                                                 $1.671\pm
                                                                 0.090$
                                                          & $2.180\pm
                                                            0.096$ &
                                                                    $3.48\pm
                                                                    0.10$
             & $67.3\pm 1.3$ & $-$ & $0.310\pm 0.012$ &
                                                                   $2.263\pm 0.028$ &
                                                                 $0.1413\pm
                                                                 0.0066$
  & $3.101\pm 0.081$ & $24.110\pm 0.030$ & $-0.070\pm 0.023$ &
                                                               $1090.00\pm
                                                               0.22$ &
                                                                       $1060.00\pm 0.29$ &
                                                               
                                                                     $0.094\pm
                                                                                           0.063$
                                                       & $2.0\pm 1.4$ &
                                                               694.93/747
                                               & 0.913\\[2mm]
&CS (4 knots)& $-$ & $-$ & $-$ &
                                                                 $-$
                                                          & $2.380\pm
                                                            0.091$&
                                                                   $3.782\pm
                                                                   0.094$
  & $61.98\pm 0.50$ & $-$ & $0.3657\pm 0.0065$ &
                                                             $2.262\pm 0.028$ &
                                                                 $0.1416\pm
                                                                 0.0066$
  & $3.104\pm 0.081$ & $24.050\pm 0.095$ & $-0.070\pm 0.023$ &
                                                               $1090.00\pm 0.23$ &
                                                                     $1060.00\pm
                                                                                   0.29$ &
                                                               
                                                                     $0.322\pm
                                                                                           0.078$
                                                       & $0.41\pm 0.25$ &
                                                               699.47/751
                                                                       &
                                                                         0.911\\[2mm]

\end{tabular}
}
\end{center}
\end{sidewaystable}

\begin{figure}
\begin{center}
\includegraphics[scale=0.55]{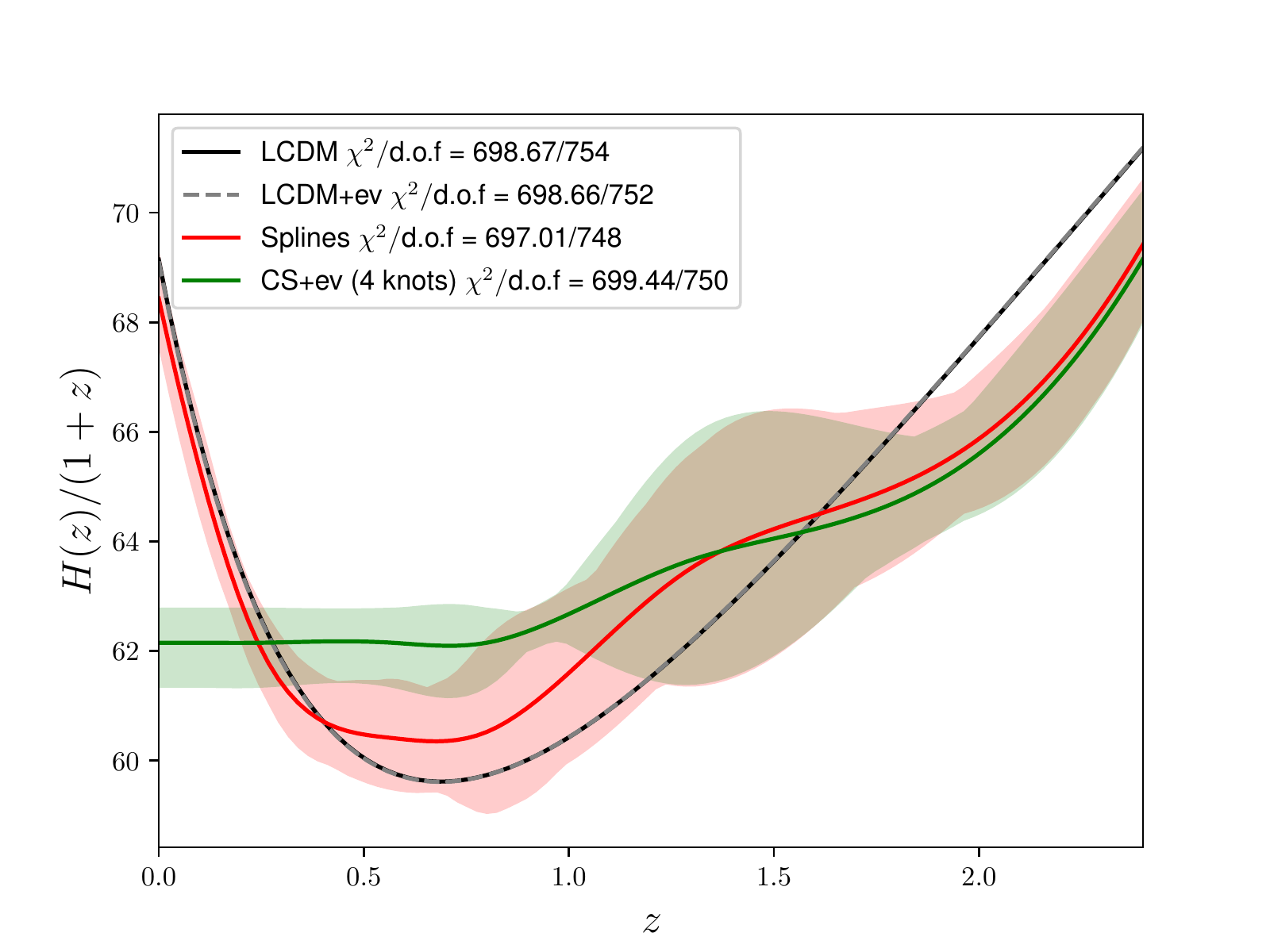}\\\includegraphics[scale=0.55]{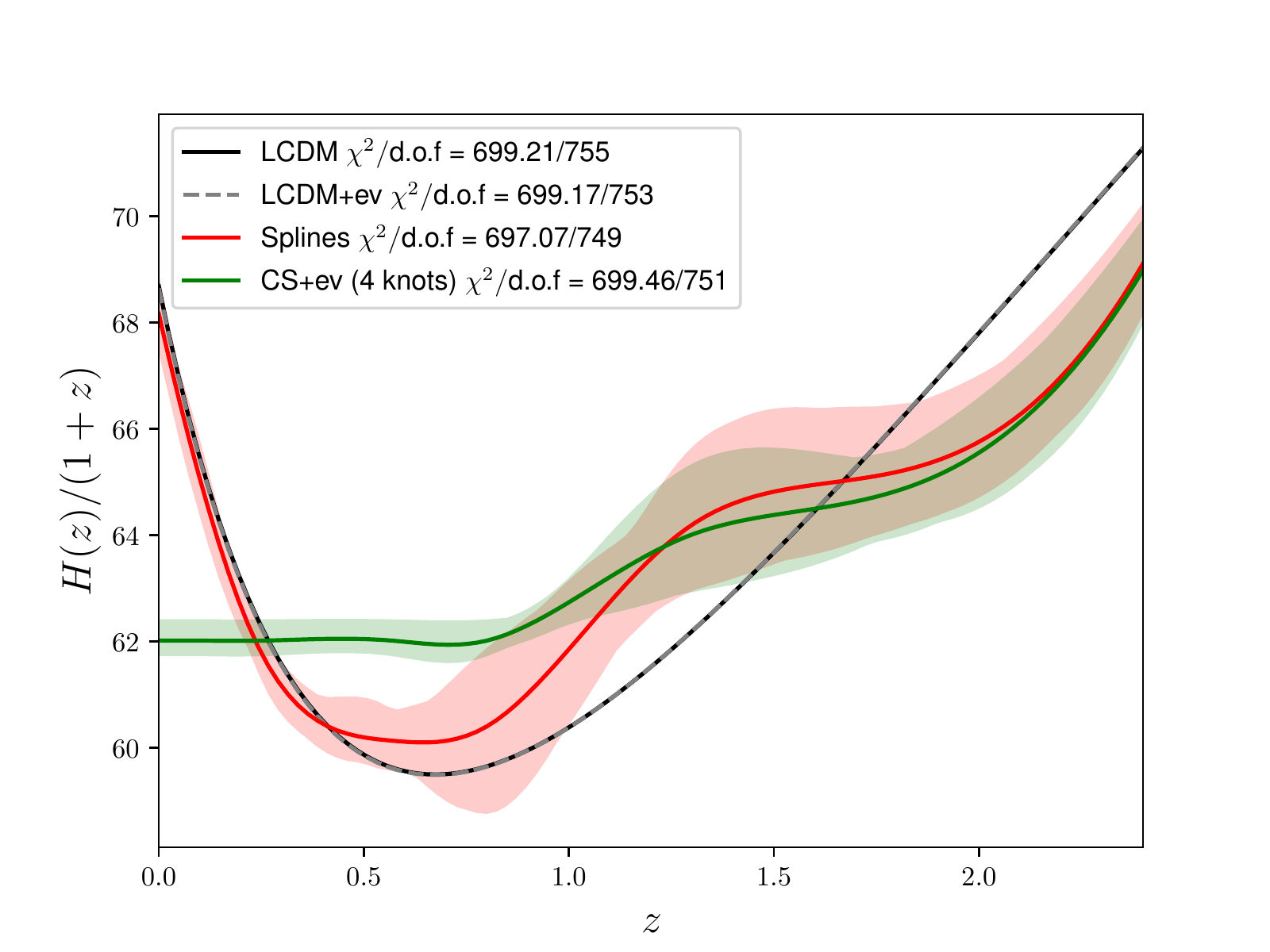}\\\includegraphics[scale=0.55]{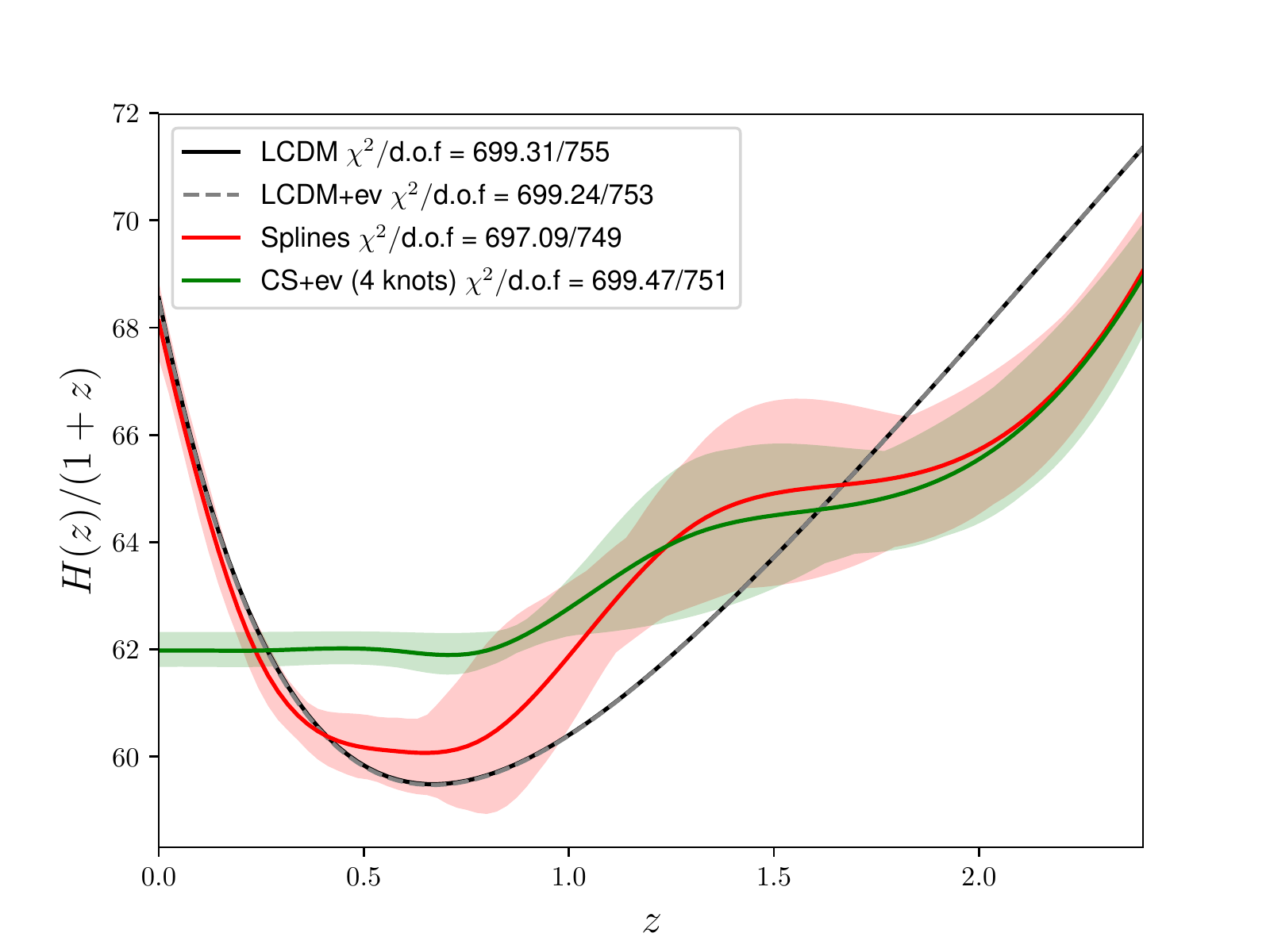}
\caption{Reconstruction of the expansion rate, $H(z)/(1+z)$, as a function of the
  redshift using the combination of SNIa, BAO, and CMB data. In the top
  panel the data sets have been combined considering $r_d$ a free
  parameter, while in the central panel a prior on $r_d$ has been
  used, and it has been explicitly computed in the bottom panel.
  In all panels the black and grey lines represent the
  $\Lambda$CDM model (without and with SNIa luminosity evolution, respectively), while the red band shows the reconstruction with
  $\Delta \chi^2\leq 1$ with respect to the best reconstruction (red
  line). The green band stands for the reconstruction of a coasting
  universe at low-redshift when SNIa intrinsic luminosity is allowed
  to vary as a function of the redshift. See the text for the details
  of the reconstruction.}\label{fig3}
\end{center}
\end{figure}

\begin{figure*}
\begin{center}
\includegraphics[scale=0.55]{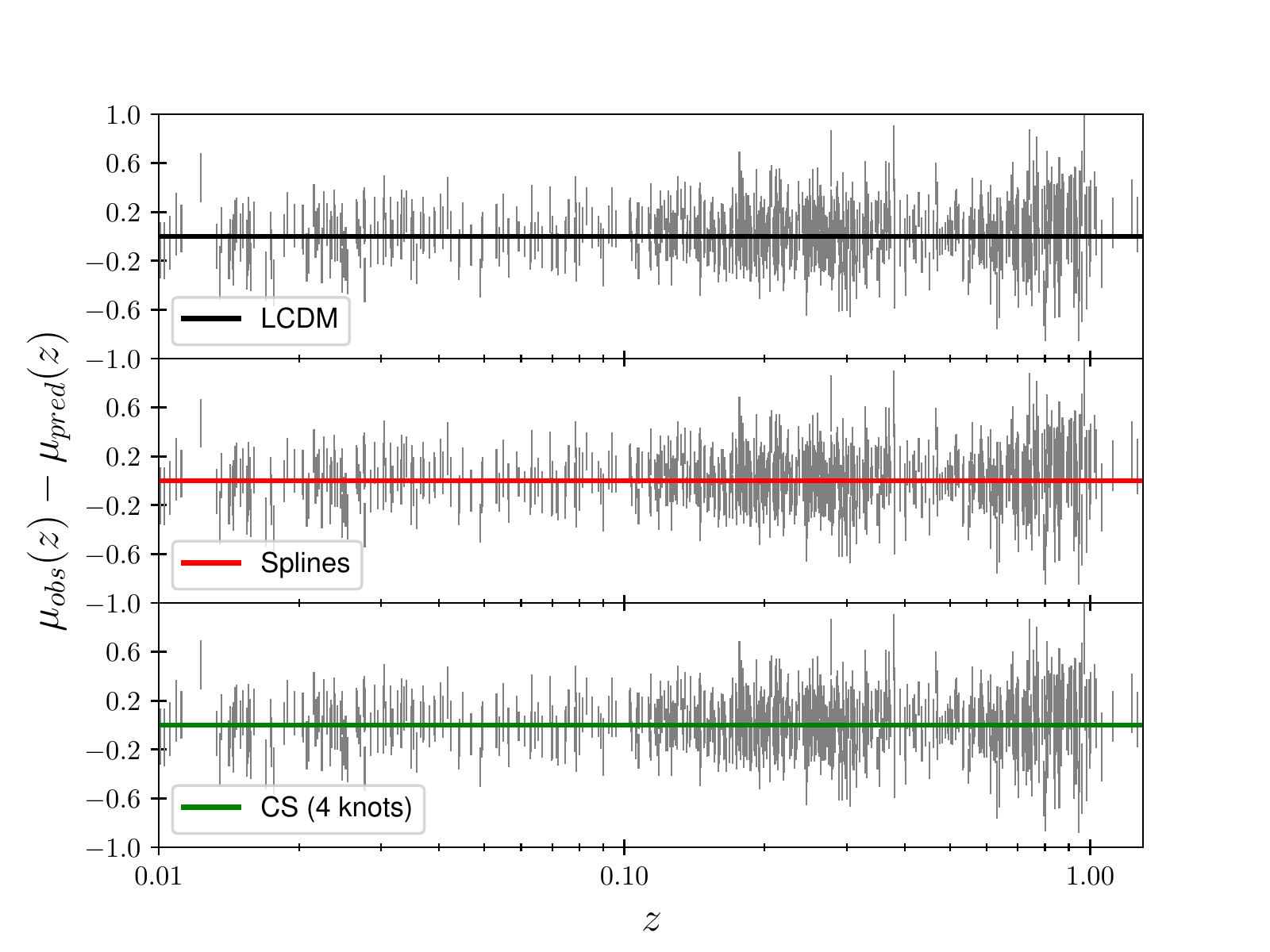}\,\includegraphics[scale=0.55]{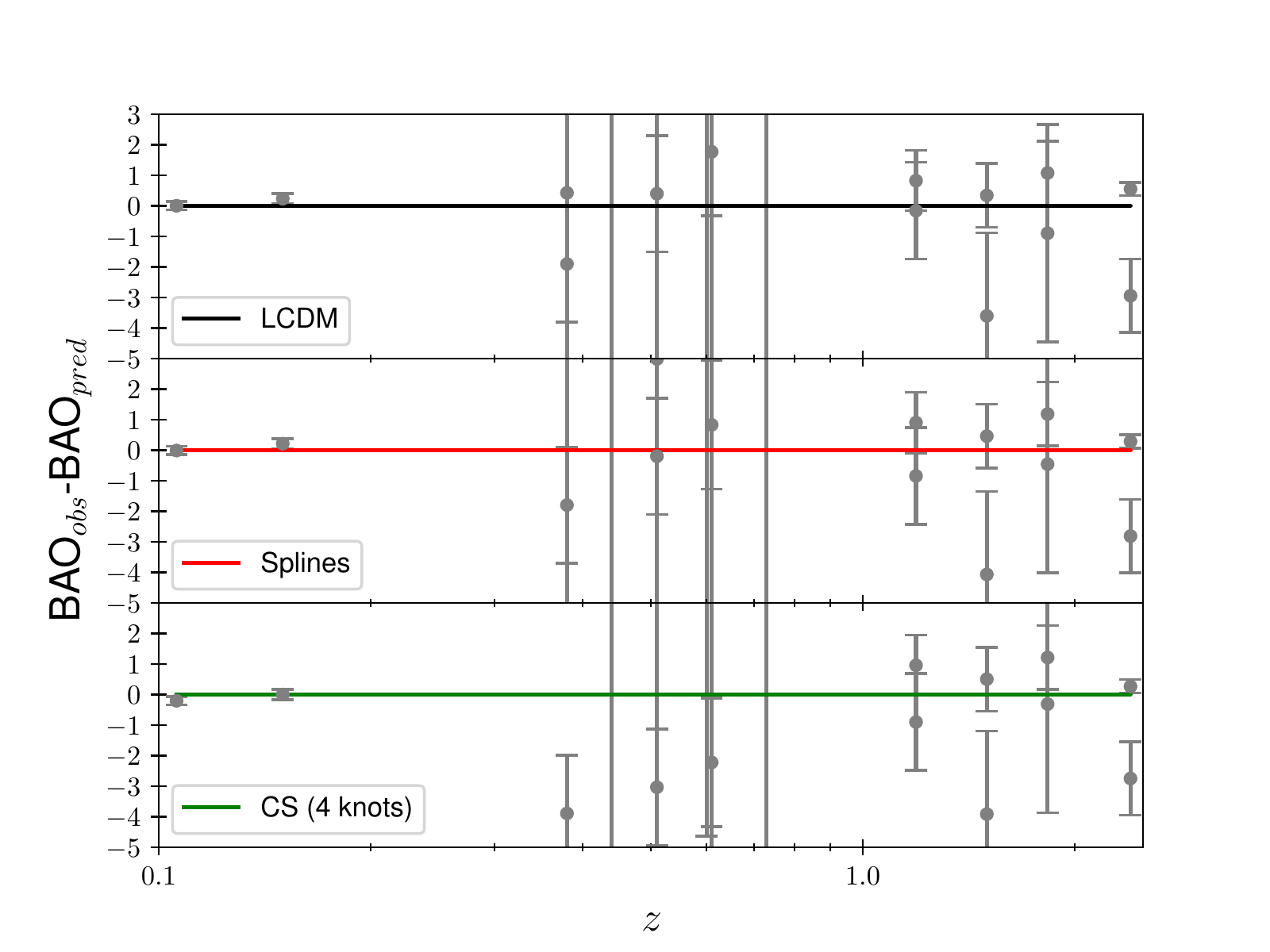}
\caption{Prediction of the different models, $\Lambda$CDM, spline
  reconstruction, and coasting reconstruction with SNIa intrinsic
  luminosity evolution, for the SNIa and BAO observables. The
  predictions have been computed using the best-fit values for the
  parameters obtained from the fit of the combination SNIa+BAO+CMB
  computing $r_d$ explicitly. {\it Left plot:} Residuals of the SNIa
  distance modulus for the three different models: $\Lambda$CDM (black
top panel), spline reconstruction (red central panel), and coasting
reconstruction (green bottom panel). The residuals have been
normalized with respect to the prediction for each model. {\it Right
  plot:} Residuals of the BAO measurements following the same colour
convention as in the left panel. The residuals have been normalized
with respect to the prediction for each model.}\label{fig4}
\end{center}
\end{figure*}

\begin{table}
  \caption{Prediction of the different models for the CMB quantities
    $R$, $\ell_a$, $\omega_b$, for the combination of SNIa, BAO, and CMB data computing $r_d$
    explicitly, and accounting for SNIa intrinsic luminosity evolution
  as a function of the redshift when dealing with a coasting reconstruction. The observed values are added as a reference.}\label{table5}
\begin{center}
\resizebox{\columnwidth}{!}{
\begin{tabular}[c]{lllll}
&Observed & $\Lambda$CDM & Splines & CS (4 knots)+ev \\[2mm]\hline \hline 
$R$ & $1.7382\pm 0.0088$ & 1.7414 & 1.7385 & 1.7382\\[2mm]
$\ell_a$ & $301.63\pm 0.15$ & 301.68 & 301.67 & 301.65\\[2mm]
$10^2\omega_b$ & $2.262\pm 0.029$ & 2.254 & 2.261 & 2.262 

\end{tabular}
}
\end{center}
\end{table}

As a last case we consider the combination of the three main
background expansion cosmological probes: SNIa, BAO, and CMB. We have
already presented two different ways to combine SNIa and BAO data, so
when we add CMB data we keep this approach and, since we now include
the physics of the early Universe, we add a third way
consisting on computing the explicit value of $r_d$. The best-fit
values, with the 1$\sigma$ errors, for the parameters for these three
subcases are presented in Table \ref{table4}, and the corresponding
reconstruction in Fig. \ref{fig3}.

Let us start with the combination considering $r_d$ a free
parameter. Both assuming the SNIa intrinsic luminosity to be redshift
independent or allowing it to vary, the three models provide
compatible values for all the parameters except $H_0$, which is
significantly smaller for the coasting reconstruction, as it was
already shown in the combination of SNIa and BAO data, and which is
compensated by a larger $\Omega_m$. When SNIa
luminosity is allowed to vary, a coasting reconstruction provides
roughly the same $\chi^2$ (699.44) as $\Lambda$CDM (698.66) with a
slightly smaller probability (0.906 against 0.918), showing that a
non-accelerated expanding universe can fit the three main background
probes when SNIa intrinsic luminosity is allowed to vary.

In a second place we add a prior on $r_d$. All the best-fit values are
compatible between the different models as before, except for $H_0$
and $\Omega_m$, which are smaller and larger for a coasting
reconstruction, respectively, accounting for SNIa luminosity evolution
or not. The obtained $\chi^2$ values are very similar, leading to very
similar probabilities to correctly fit the data, and they show that a
coasting reconstruction can correctly fit the data when SNIa
luminosity evolution is accounted for. In the last place we compute
$r_d$ using equation (\ref{eqrd}). All the best-fit values are
compatible with the previous results, and compatible between the
different models, except for $H_0$ and $\Omega_m$. It is also the case
for the $\chi^2$ values and the corresponding probabilities. We conclude that a non-accelerated universe can correctly fit the three main
background probes when we account for a redshift dependence in the
intrinsic luminosity of SNIa. 

For completeness, we present in
Fig. \ref{fig4} the residuals to SNIa and BAO observations for three
different models: $\Lambda$CDM (black), the reconstruction through
cubic splines (red), and the non-accelerated model using a coasting
reconstruction (green) taking into account SNIa intrinsic luminosity
evolution. We also provide the predictions for the CMB quantities $R$,
$\ell_a$, and $\omega_b$ in Table \ref{table5}. All these predictions
have been computed using the best-fit values for the parameters
obtained from the global fit to the combination of SNIa, BAO, and CMB
data, computing explicitly the value of $r_d$ using equation
(\ref{eqrd}). From these results we can graphically see that all three
models are perfectly able to fit the data; including the coasting
reconstruction with SNIa luminosity evolution. As it can be seen in
Table \ref{table4}, a different approach when combining SNIa, BAO, and
CMB data gives nearly the same values for the parameters, which leads
to nearly the same predictions.

\subsection{Growth rate}
\begin{figure}
\begin{center}
\includegraphics[scale=0.55]{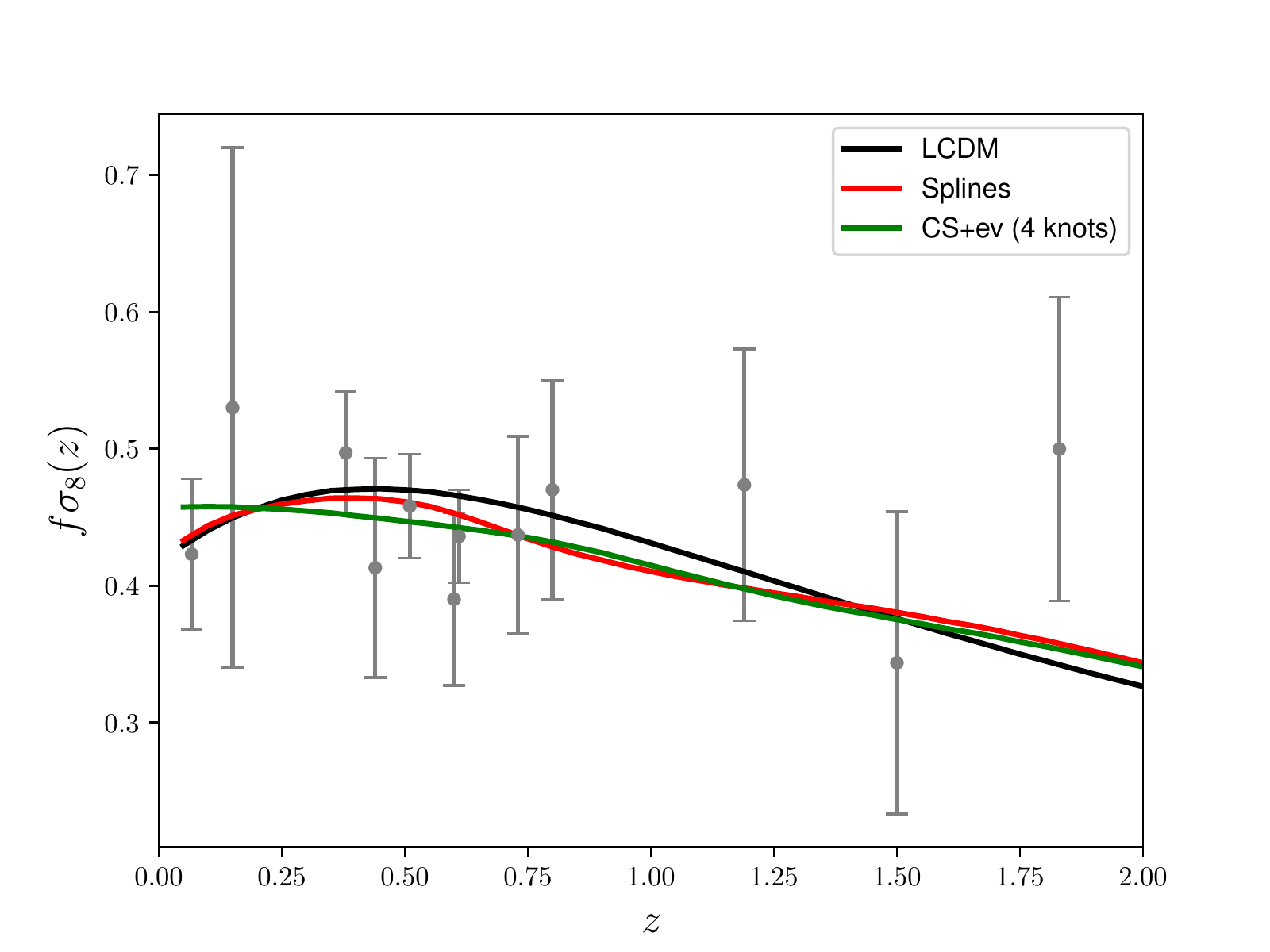}
\caption{Prediction of the different models, $\Lambda$CDM, spline
  reconstruction, and coasting reconstruction with SNIa intrinsic
  luminosity evolution, for the growth of matter perturbations $f\sigma_8$
observable. The predictions have been computed using the best-fit
values for the parameters obtained from the fit of the combination
SNIa+BAO+CMB computing $r_d$ explicitly. Therefore, it is not a fit to the
$f\sigma_8$ measurements. We follow the same colour legend as in the
previous figures: black for $\Lambda$CDM, red for the spline
reconstruction, and green for the coasting
reconstruction.}\label{fig5}
\end{center}
\end{figure}

The measurements of the growth rate of matter perturbations offer an
additional constraint on cosmological models. Their value depend on
the theory of gravity used and it is well known that there are identical
background evolutions with different growth rates
[\cite{fsig8_1}]. Defining the linear growth factor of matter
perturbations as the ratio between the linear density perturbation and
the energy density, $D\equiv \delta \rho_m/\rho_m$, we can derive the
standard second order differential equation for $D$ [\cite{fsig8_2}]

\begin{equation}
\ddot{D}+2H\dot{D}-4\pi G\rho_mD=0\,,
\end{equation}
where the dot stands for differentiation over the cosmic
time. Neglecting second order corrections, this differential equation
can be rewritten with derivatives over the scale factor [\cite{fsig8_3}]

\begin{equation}
D''(a)+\left[\frac{3}{a}+\frac{H'(a)}{H(a)}\right]D'(a)-\frac{3}{2}\Omega_m\frac{H_0^2}{H^2(a)}\frac{D(a)}{a^5}=0\,,
\end{equation}
which is valid under the assumption that dark energy cannot be
perturbed and does not interact with dark matter. We can now define
the growth factor as

\begin{equation}
f(a)\equiv \frac{\text{d ln} D}{\text{d ln} a}\,,
\end{equation}
and then compute the observable weighted growth rate $f\sigma_8$ as

\begin{equation}
f\sigma_8(z)=f(z)\cdot\left(\sigma_{8_{\rm Planck}}\frac{D(z)}{D_{\rm Planck}(0)}\right)\,,
\end{equation}
where $\sigma_{8_{\rm Planck}}$ stands for the observed value of the
root mean square mass fluctuation amplitude on scales of $8h^{-1}$ Mpc
at redshift $z=0$ (fixed to 0.8159 in this work [\cite{CMB}]), and $D_{\rm Planck}$ represents the $\Lambda$CDM
Planck growth rate [\cite{CMB}]. In this work we consider the measurements of the
weighted growth rate from the 6dFGS survey [\cite{fsig8_4}], the WiggleZ
survey [\cite{fsig8_5}], and the VIPERS survey [\cite{fsig8_6}], as well
as the different SDSS projects: SDSS-II MGS DR7 [\cite{fsig8_7}] (with
the main galaxy sample of the seventh data release), SDSS-III BOSS
DR12 [\cite{BOSSfinal}] (with the LRGs from the 12th BOSS data release),
and SDSS-IV DR14Q [\cite{eBOSS1}] (with the latest quasar sample of
eBOSS). We have not included this data set in our fitting analysis for
simplicity, but we show in Fig. \ref{fig5} that, using the best-fit
values for the parameters from the SNIa+BAO+CMB fit, the prediction
for the three models considered ($\Lambda$CDM, spline reconstruction,
and coasting reconstruction with SNIa luminosity evolution) is in very
good agreement with the observations. Notice that the values for the
parameters used in Fig. \ref{fig5} have been obtained computing the
value of $r_d$, but the results are equivalent using the other
approaches for the combination of our three main data sets.

\subsection{The Hubble constant}
\begin{figure*}
\begin{center}
\includegraphics[scale=1]{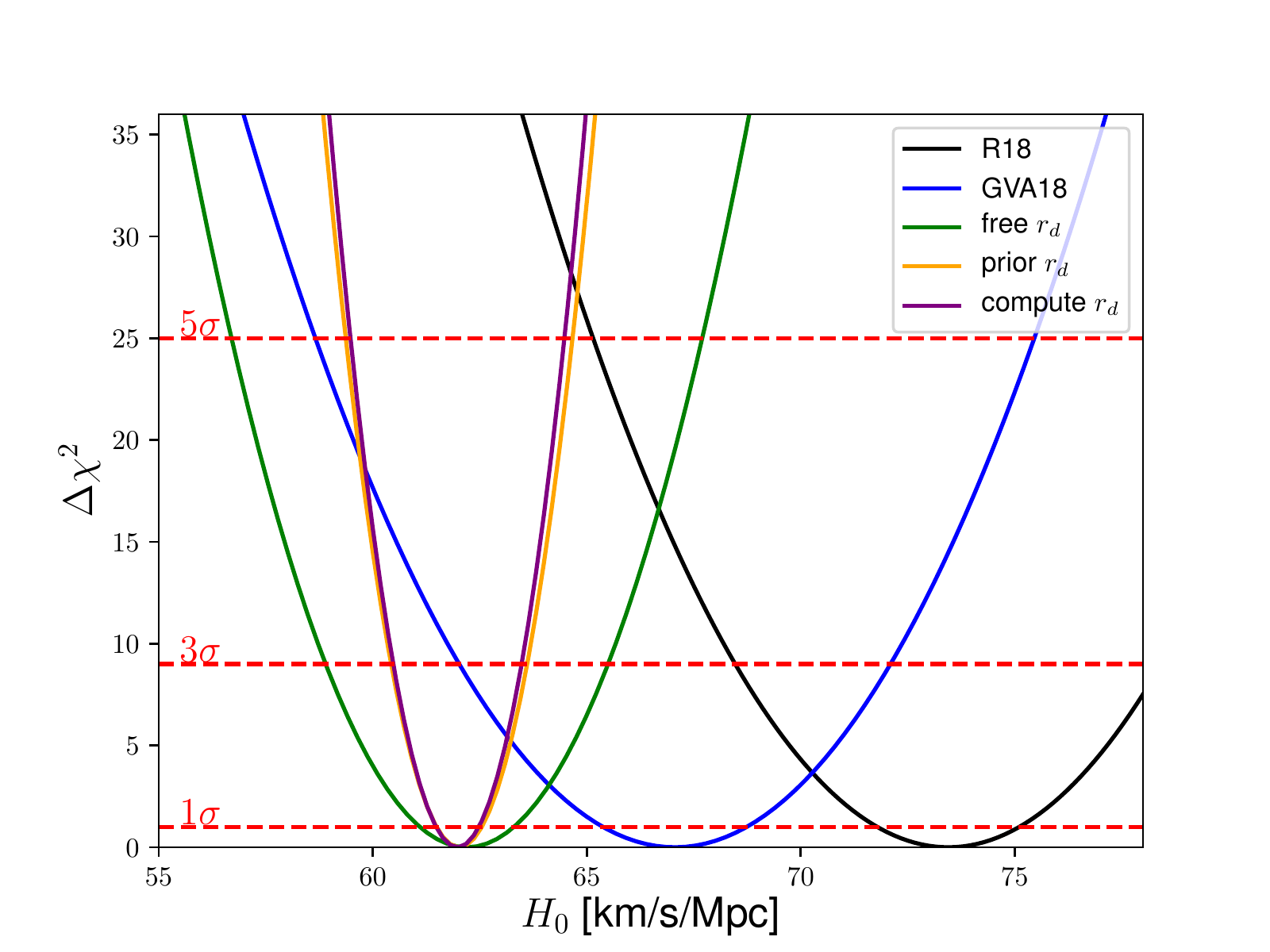}
\caption{Profile likelihood (assuming Gaussian likelihoods) of
  different values for the Hubble constant. The black line corresponds to the
  value measured from the HST (R18), while the blue one stands for the
  measured value from SNIa and $H(z)$ data using Gaussian Processes
  (GVA18). The other three profiles represent the predicted value from a
  non-accelerated reconstruction with different approaches to combine
  the three main data sets of this work (SNIa, BAO, and CMB): consider
$r_d$ a free parameter (green), add a prior on it (yellow), or compute
it explicitly (purple). The 1$\sigma$, 3$\sigma$, and 5$\sigma$ lines
are represented as a reference.}\label{fig6}
\end{center}
\end{figure*}
The Hubble constant, $H_0$, is one of the most important parameters in modern
cosmology, since it is used to construct time and distance
cosmological scales. It was first measured by Hubble to be roughly 500
km/s/Mpc [\cite{HubbleH0}]. Current data supports a value for $H_0$
close to 70 km/s/Mpc. However, nearly 100 years later there is
still no consensus on its value. Local measurements already show some
tension on the results depending on the calibration of SNIa distances
[\cite{H0local1,H0local2}]. Moreover, there is also some tension
between the direct measurement of $H_0$ and the value inferred from
the CMB assuming a $\Lambda$CDM model [\cite{CMB}]. There has been
many attempts in the literature to solve this discrepancy both from an
observational and a theoretical perspective [see \cite{Bernal,GV}
and references therein for a detailed discussion on the trouble with
$H_0$]. In this work we consider two very recent model
independent measurements of $H_0$ in order to check its effect on the
conclusions we can draw concerning the cosmic acceleration. We first consider the value obtained from the Hubble Space
Telescope observations in \cite{H0local1}
(R18 in the following), $H_0=73.45\pm 1.66$ km/s/Mpc. We then consider
the value
obtained with Gaussian Processes using SNIa data, and constraints on
$H(z)$ from cosmic chronometers in \cite{GV} (GVA18 in the following),
$H_0=67.06\pm 1.68$ km/s/Mpc. This last value is closer to the one
derived with an ``inverse distance ladder'' approach in
\cite{AubourgH0}, $H_0=67.3\pm 1.1$, where the measurement assumes standard pre-recombination
physics but is insensitive to dark energy or space
curvature assumptions. It is also closer to the value derived from the
CMB observations using a flat $\Lambda$CDM model, $H_0=67.51\pm 0.64$ [\cite{CMB}].

In Fig. \ref{fig6} we represent the profile likelihood (assuming
Gaussian likelihoods) for both the observed values of $H_0$, R18 (black) and
GVA18 (blue), and the values derived from the non-accelerated reconstruction for
the combination SNIa+BAO+CMB taking into account the SNIa intrinsic
luminosity evolution. We present the three values obtained for the
three approaches followed when combining the data sets: consider $r_d$
a free parameter (green), add a prior on it (yellow), or compute it
explicitly (purple). From the figure alone it is clear that the $H_0$
value for the non-accelerated reconstruction is in tension with R18 at
more than 5$\sigma$, independently of the approach used when combining
the data sets. More precisely, a non-accelerated reconstruction is
ruled out if we consider the R18 measurement at 5.65$\sigma$ (free
$r_d$, $H_0=62.2\pm 1.1$), 6.56$\sigma$ (prior $r_d$, $H_0=62.02\pm
0.53$), or 6.62$\sigma$ (compute $r_d$, $H_0=61.98\pm
0.50$),
showing that, with the R18 measurement, cosmic acceleration is proven
even if some astrophysical systematics evolving with the redshift
modify the intrinsic luminosity of SNIa. However, we can also see from
the figure that if we consider the measured value from the Gaussian
Processes, a non-accelerated reconstruction shows a bit less than a
3$\sigma$ tension. More precisely, there is a tension of 2.42$\sigma$
(free $r_d$), 2.86$\sigma$ (prior $r_d$), or 2.90$\sigma$ (compute
$r_d$). In this case, the measured value of $H_0$ points towards
ruling out these reconstructions, but the tension is still far from
the 5$\sigma$ threshold.

\section{Conclusions}\label{sec5}
In this paper we have adressed the question whether relaxing the standard
assumption that SNIa intrinsic luminosity does not depend on the
redshift may have an impact on the conclusions that can be drawn on
the accelerated nature of the expansion of the
Universe. Although there is no theoretically motivated model for this
luminosity evolution, it has not been proven that two SNIa in two galaxies with the
same light-curve, colour, and host stellar mass have the same
intrinsic luminosity independently of the redshift. Moreover, with this kind of analysis we can distinguish between the
effect of unknown astrophysical systematics varying with the redshift
and the cosmological information.

The impact of SNIa luminosity evolution on our cosmological knowledge
has already been adressed before [\cite{Wright,Drell,
    SNIaevII, Nordin, Ferr2009,Tutusaus,Tutusaus2}],
but in this work we have extended the analysis by including the
physics of the early Universe ($z\approx 1000$), and thus
considering the main background
cosmological probes: SNIa, BAO, and the CMB. In order to be as general as
possible, we have not imposed a cosmological model, but we have reconstructed
the expansion rate of the Universe using a cubic spline
interpolation. 

We have first
applied, as an illustration of the method, the reconstruction to SNIa data
alone with the standard SNIa luminosity independence assumption. We
have shown that with this assumption cosmic acceleration is definitely
preferred against a local non-accelerated universe.

In a second step we have added the latest BAO data to our analysis. We
have considered two different ways to combine it with SNIa data:
either we have considered $H_0r_d$ as a free parameter, or we have
added a prior on $r_d$ coming from CMB observations, without any dependence on late-time Universe
assumptions. In both cases we have seen that a non-accelerated
universe is able to fit the data nearly as nicely as $\Lambda$CDM,
when we allow the SNIa intrinsic luminosity to vary as a function of
the redshift.

Next, we have extended the data sets in the analysis by adding the
information coming from the CMB through the reduced parameters. In
order to deal with this information we have been forced to specify
the model up to very high redshifts. We have decided to follow a
matter dominated model (plus radiation and a negligible dark energy
contribution in the form of a cosmological constant) from the early
Universe down to $z\approx 3$,
where we start to have low-redshift data. We have then coupled the
model to our spline reconstruction. In other words, we consider a
matter-radiation dominated model at the early Universe and, when we
start to have low-redshift data and a cosmological constant is still negligible, we allow the expansion rate to vary freely without
specifying any dark energy model. When adding
the CMB data we follow three different approaches: treat $r_d$ as a
free parameter, add a prior on it, or compute it assuming that the BAO
and the CMB share the same physics. In all three cases we have seen
that a non-accelerated model is able to nicely fit the data, when SNIa
intrinsic luminosity is allowed to vary, including the information on
the early Universe coming from the CMB. 

For simplicity we have not added the $f\sigma_8$ measurements for the
growth rate of matter perturbations, but we have checked that using
the best-fit values from the global fit SNIa+BAO+CMB we are able to
correctly predict the latest $f\sigma_8$ measurements.

After having seen that if SNIa intrinsic luminosity does depend on the
redshift, the main cosmological probes are not able to rule out a
non-accelerated model, we focus on the impact that the Hubble constant
may have on this question. We have considered two different model
independent recent measurements of $H_0$: $73.45\pm 1.66$ km/s/Mpc
(R18) from \cite{H0local1}, and $67.06\pm 1.68$ km/s/Mpc (GVA18) from
\cite{GV}. We have shown that if we consider the R18 value, cosmic
acceleration is proven at more than 5.65$\sigma$ for a general
expansion rate reconstruction [for which we get $H_0=62.2\pm 1.1$
(free $r_d$), $H_0=62.02\pm 0.53$ (prior $r_d$), and $H_0=61.98\pm
0.50$ (compute $r_d$)], even if SNIa intrinsic luminosity
varies as a function of the redshift due to any astrophysical unknown
systematic. It is important to say, though, that if we consider the GVA18
value, a non-accelerated reconstruction for the expansion rate is at a
3$\sigma$ tension with the measurement, but still below the 5$\sigma$ detection. 

In conclusion, if SNIa intrinsic luminosity varies as a function of
the redshift, a non-accelerated universe is able to correctly fit all
the main background probes. However, the value of $H_0$ turns out to
be a key ingredient in the conclusions we can draw concerning the cosmic acceleration. If we take it into
account we are close to claim an accelerated expansion of the Universe
using an approach very independent of the cosmological model assumed,
and even if SNIa intrinsic luminosity varies. A final consensus on a
direct measurement of $H_0$ and its precision will be decisive to
finally prove the cosmic acceleration independently of the
cosmological model and any redshift
dependent astrophysical systematic that may remain in the SNIa analysis.

\begin{acknowledgements}
We thank Adam G. Riess and Daniel L. Shafer for very fruitful comments
which helped to noticeably improve this work.

This work has been carried out thanks to the support of the OCEVU Labex (ANR-11-LABX-0060) and of the Excellence Initiative of Aix-Marseille University - A*MIDEX, part of the French “Investissements d’Avenir” programme.
\end{acknowledgements}

%
\bibliographystyle{aa} 
\bibliography{aa} 
%

\end{document}